\documentclass[12pt]{article}
\pdfoutput=1
\usepackage[nosort]{cite}
\usepackage{amsmath,amssymb}

\usepackage{color}

\usepackage{cite}
\usepackage{hyperref}
\usepackage{tikz}
\usepackage{slashed}
\usepackage{subcaption}

\usepackage{draft}

\def\({\left(}
\def\){\right)}
\newcommand{\beq}{\begin{equation}\begin{aligned}}
\newcommand{\eeq}{\end{aligned}\end{equation}}
\def\d{\partial}

\begin{document}

\begin{titlepage}
	\hfill MIT-CTP/5336
	\begin{center}
		
		\hfill \\
		\hfill \\

		\title{Anomaly Inflow for Subsystem Symmetries}
		
		\author{Fiona J. Burnell$^{1,2}$, Trithep Devakul$^{3,4}$, Pranay Gorantla$^{4}$, Ho Tat Lam$^{3,4}$, \\and Shu-Heng Shao$^{2,5}$}
		
		\address{${}^{1}$School of Physics and Astronomy, University of Minnesota, Minneapolis MN, USA\\
		${}^{2}$School of Natural Sciences, Institute for Advanced Study, Princeton NJ, USA
		\\
		${}^{3}$Department of Physics, Massachusetts Institute of Technology, Cambridge MA, USA
		\\
		${}^{4}$Physics Department, Princeton University, Princeton NJ, USA
		\\
		${}^{5}$C.\ N.\ Yang Institute for Theoretical Physics, Stony Brook University, Stony Brook NY, USA}
	\end{center}
	
	\vspace{2.0cm}
	
	\begin{abstract}
		
		\noindent

		We study 't Hooft anomalies and the related anomaly inflow for subsystem global symmetries.  
		These symmetries and anomalies arise in a number of exotic systems, including models with fracton order such as the X-cube model.  
		As is the case for ordinary global symmetries, anomalies for subsystem symmetries  can be canceled by anomaly inflow from a bulk theory in one higher dimension; the corresponding bulk is therefore a non-trivial subsystem symmetry protected topological (SSPT) phase. 
		We demonstrate these phenomena in several examples with continuous and discrete subsystem global symmetries, as well as time-reversal symmetry. 
		For each example we describe the boundary anomaly, and present classical continuum actions for the corresponding bulk SSPT phases, which describe the response of background gauge fields associated with the subsystem symmetries.  
		Interestingly, we show that the anomaly does not uniquely specify the bulk SSPT phase. In general, the latter may also depend  
 on how the symmetry and the associated foliation structure on the boundary are extended into the bulk.
 		
		\end{abstract}
	
	\vfill
	
\end{titlepage}

\eject

\tableofcontents

\section{Introduction}

Global symmetry is one of the central tools in analyzing strongly-coupled quantum systems.  
In recent years, a new kind of global symmetry, known as the \textit{subsystem global symmetry}, has featured prominently in many exotic lattice systems, including the gapless model of \cite{PhysRevB.66.054526} and many gapped fracton models \cite{PhysRevLett.94.040402,PhysRevA.83.042330}.\footnote{Subsystem global symmetries have also appeared in some earlier references  such as \cite{PhysRevLett.85.2160}.} 
(See \cite{Nandkishore:2018sel,Pretko:2020cko} for reviews on fractons.) 
In this paper, we will discuss  anomaly inflow \cite{Callan:1984sa} and symmetry-protected topological (SPT) phases for subsystem global symmetries \cite{Raussendorf2019,You2018,Devakul2018,Stephen2019subsystem,You2020,Devakul2020}.  
  We will  be working under the framework developed in \cite{Seiberg:2019vrp,paper1,paper2,paper3,Gorantla:2020xap,Gorantla:2020jpy,Rudelius:2020kta,Gorantla:2021svj,Gorantla:2021bda} for these exotic field theories with subsystem global symmetries.

Unlike for ordinary global symmetry, the generator of a subsystem global symmetry acts only on a subspace $\mathcal{S}$ of the whole spatial manifold.\footnote{Similar to the term higher-form global symmetry, the adjective ``global" does not mean that the charges act globally on the whole space. Rather, it is used to distinguish the case of interest from that of  the gauge symmetry.}  
Different  choices of the subspace generally give rise to independent conserved charges. 
On a lattice, the number of independent conserved charges therefore grow sub-extensively with the number of sites. 
In the low-energy limit, this leads to an infinite number of charges, which underlies many of the peculiarities in these exotic models. 
These include the surprising UV/IR mixing in some of the physical observables  \cite{paper1,paper2,paper3,Gorantla:2020xap,Slagle:2020ugk,Gorantla:2020jpy,Rudelius:2020kta, You:2021tmm,Gorantla:2021svj,Zhou:2021wsv,Hsin:2021mjn,You:2021sou,Gorantla:2021bda}.

It is useful to compare the subsystem global symmetry with another generalized symmetry, the higher-form global symmetry \cite{Gaiotto:2014kfa}. 
For both kinds of global symmetries, the conserved charges are supported on closed,  higher-codimensional manifolds $\mathcal{S}$ in space.  
But the charges, especially in the continuum limit, are different in many ways for these two kinds of symmetries. 
The charge $Q(\mathcal{S})$ of a  higher-form symmetry depends on $\mathcal{S}$ topologically, i.e., $Q(\mathcal{S}) =Q(\mathcal{S}')$ if $\mathcal{S}$ and $\mathcal{S}'$ are homologous to each other. 
Relatedly, there is no restriction on the choice of the manifold $\mathcal{S}$ of a given codimension.  
On the other hand, the charge of a subsystem global symmetry depends not only on the topology of $\mathcal{S}$, but possibly also on the shape and the location of $\mathcal{S}$.  Furthermore, the charge might only  be allowed to be on certain $\mathcal{S}$ but not all  manifolds of a given codimension. (For example, $\mathcal{S}$ may be restricted to be straight lines along certain preferred directions, rather than be a general curve.)  
See \cite{Seiberg:2019vrp,Qi:2020jrf} for related discussions.

Just as for ordinary global symmetry, one can attempt to gauge a subsystem global symmetry by coupling to dynamical gauge fields.  
This is, however, not always possible. 
The obstruction to gauging a global symmetry is known as the \textit{'t Hooft anomaly}.

\begin{center}
	\emph{Review of anomaly inflow}
\end{center}

The 't Hooft anomaly of a quantum system $\cal T$ in $d$ spacetime dimensions, with either ordinary or subsystem global symmetry, can be diagnosed as follows. 
We couple the system $\cal T$ to  background gauge fields $A$ and denote the partition function by $Z[A]$.  
For an ordinary   global symmetry,  $A$ are  one-form gauge fields. 
For a subsystem global symmetry,  they are tensor gauge fields, as we describe in detail in the main text. 
When an 't Hooft anomaly is present, under a background gauge transformation $A\to A^g$, the partition function is not invariant but transforms with an anomalous phase:
\ie
Z_{\cal T}[A^g]  = e^{ i \int_{M^{(d)}} \theta(g,A) } Z_{\cal T}[A]\,,
\fe
where $M^{(d)}$ is the spacetime $d$-dimensional manifold. 
We can always {\it change} the anomalous phase $\theta(g,A)$ by adding $d$-dimensional  local counterterms of the background gauge fields $A$. 
However, 
the 't Hooft anomaly is characterized by the fact that no choice of $d$-dimensional local counterterms can remove the anomalous phase.  

Another powerful way to describe the anomaly is using a classical field theory   in one dimension higher. 
This classical field theory is the continuum description of the SPT phase. 
Let the partition function for this  classical field theory of the background gauge fields $A$ in $d+1$ spacetime dimensions be 
\ie
\exp\left( i \int_{N^{(d+1)}} \omega(A)\right)\,.
\fe 
If $N^{(d+1)}$ has no boundary, then this partition function is gauge invariant. 
When $N^{(d+1)}$ has a boundary, then there can be  a boundary term under the background gauge transformation. 
Let $N^{(d+1)}$ by a $d+1$-dimensional manifold whose boundary is $M^{(d)}$, i.e., $\partial N^{(d+1)}  = M^{(d)}$. 
While a genuine anomaly of $\cal T$  cannot be canceled by a $d$-dimensional local counterterm, it can generally be canceled by the anomalous gauge transformation of a classical field theory in $d+1$ spacetime dimensions:
\ie
\exp\left( i \int_{N^{(d+1)}} \omega(A^g )\right) = \exp \left(-i \int_{M^{(d)} } \theta(g,A)\right) \exp\left( i \int_{N^{(d+1)}} \omega(A)\right)\,.
\fe  
In other words, the original $d$-dimensional system $\cal T$ coupled to a $d+1$-dimensional bulk classical field theory 
\ie
Z_{\cal T}[A] \exp\left( i \int_{N^{(d+1)}} \omega(A)\right)
\fe
 is  invariant under the background gauge transformation.

We emphasize that there is nothing inconsistent with the original system $\cal T$ in $d$ spacetime dimensions with an anomalous (subsystem) global symmetry.  Such a system can be defined without the  need of a bulk in one dimension higher.  
We simply cannot gauge the global symmetry in $d$ spacetime dimensions.

\begin{center}
	\emph{Anomaly inflow for subsystem symmetries}
\end{center}

  In this paper, we will discuss several examples of anomaly inflow for subsystem global symmetries and the corresponding subsystem symmetry-protected topological phases (SSPT) in  one dimension higher.\footnote{For the rest of this paper, we will use SSPT and the classical field theory interchangeably.}  
  Our examples include discrete and continuous subsystem symmetries, and for each one of them we will also discuss an analogous system with an anomalous ordinary  global symmetry in the appendices.

The simplest example  of an 't Hooft anomaly in a model with continuous subsystem symmetry is the  $U(1)\times U(1)$ anomaly  of the 2+1d continuum field theory of \cite{paper1}, which had been first introduced in \cite{PhysRevB.66.054526}:
   \ie
{\cal L}={\mu_0\over2} (\partial_\tau\phi)^2+{1\over 2\mu} (\partial_x\partial_y\phi)^2\,,~~~~\phi\sim \phi+2\pi\,.
\fe
(See also \cite{Tay_2011,You:2019cvs,You:2019bvu,Karch:2020yuy,You:2021tmm,Gorantla:2021svj,Gorantla:2021bda,Lake:2021pdn}  for related discussions on this theory.) 
  This anomaly, both in the continuum and on the lattice,  was discussed in \cite{Gorantla:2021svj,Gorantla:2021bda}. 
  Here we will further  present its SSPT in 3+1 dimensions, which can be described as a Euclidean Lagrangian of the  classical bulk tensor gauge fields $(A_\tau,A_{xy},A_z), (\tilde A_\tau ,\tilde A_{xy} ,\tilde A_z)$: 
  \ie
{i\over 2\pi} \left[
\tilde A_{xy} (\partial_\tau A_z - \partial_z A_\tau ) +\tilde A_z (\partial_\tau A_{xy}  -\partial_x \partial_y A_\tau)
- \tilde A_\tau (\partial_z A_{xy} - \partial_x \partial_y A_z)
\right]~.
\fe
We will discuss this 3+1d SSPT, and the associated tensor gauge fields, in more detail in Section \ref{sec:U1U1SSPT}.
Interestingly, this anomaly can be viewed as a higher-rank analog of a mixed anomaly between the momentum $U(1)$ and the winding $U(1)$ symmetry in the ordinary 1+1d compact boson.

In this work, we also analyze a number of systems with anomalies in their subsystem symmetries and SSPTs  that have not been previously discussed in the literature. 
A noteworthy example is the 3+1d $\mathbb{Z}_N$ X-cube model \cite{Vijay:2016phm}, one of the simplest gapped fracton models.  
The X-cube model has two sets of $\mathbb{Z}_N$ subsystem global symmetries, supported on strips and lines, respectively.\footnote{Many gapped fracton models arise as the gauge theory of a subsystem  symmetry \cite{Vijay:2016phm,Williamson:2016jiq,Shirley:2018vtc}. Here we discuss the subsystem global symmetry of the X-cube model, not the gauge symmetry.} On the lattice, these symmetries are simply the logical operators that map between different ground state sectors.   In the continuum field theory, they become the Wilson operators of the underlying tensor gauge fields \cite{Slagle:2017wrc,paper3}.     
We show that these two subsystem symmetries have a mixed 't Hooft anomaly, which we describe explicitly using the field theory developed in \cite{Slagle:2017wrc,paper3} (see also \cite{Slagle:2018swq,Slagle:2020ugk,Hsin:2021mjn}).
An immediate consequence of this anomaly is that the two $\mathbb{Z}_N$ subsystem symmetry operators do not commute with each other, leading to the sub-extensive ground state degeneracy \cite{paper3}. Moreover, we identify a 4+1-dimensional SSPT that cancels the anomaly of these subsystem global symmetries in the X-cube model. 

An analogy can be drawn between the X-cube model and the 2+1-dimensional toric code \cite{Kitaev:1997wr}. 
The toric code has two $\mathbb{Z}_N$ string-like logical operators which act within the space of ground states. 
They become the $\mathbb{Z}_N\times \mathbb{Z}_N$ one-form global symmetry of the 2+1-dimensional $\mathbb{Z}_N$ gauge theory, the continuum description of the toric code.  
The nontrivial commutation relation between the two $\mathbb{Z}_N$ operators can be interpreted as a mixed 't Hooft anomaly between the two $\mathbb{Z}_N$ one-form global symmetries \cite{Gaiotto:2014kfa,Gomis:2017ixy,Hsin:2018vcg}.
(See also \cite{Wen:2018zux} for  a parallel discussion  from the condensed matter viewpoint.) 
This anomaly can be canceled by a 3+1-dimensional SPT \cite{Kapustin:2013uxa,Gaiotto:2014kfa,Thorngren:2015gtw,Hsin:2018vcg}, which is the low-energy limit of a Walker-Wang model \cite{Walker:2011mda}.\footnote{More specifically, this is a  Walker-Wang model whose input braided tensor category is modular.  In this case, the low-energy limit is invertible and has no bulk topological order.  } 
Our  4+1-dimensional SSPT for the X-cube model is analogous to this ordinary 3+1-dimensional SPT.

In all of our examples, the subsystem symmetries are associated with a foliation structure in space. 
The foliation is typically specified by leaves defined by setting one of the  spatial coordinates to be a constant. 
The importance of the choice of the foliation in models with subsystem symmetries has been emphasized in \cite{Slagle:2017mzz,Shirley:2017suz,Shirley:2019wdf,Shirley:2018hkm,Shirley:2018vtc,Slagle:2018kqf,Slagle:2018swq,
Shirley:2019uou,Slagle:2020ugk,Rudelius:2020kta,Hsin:2021mjn}.

\begin{figure}
  \begin{center}
  \includegraphics[scale=0.3]{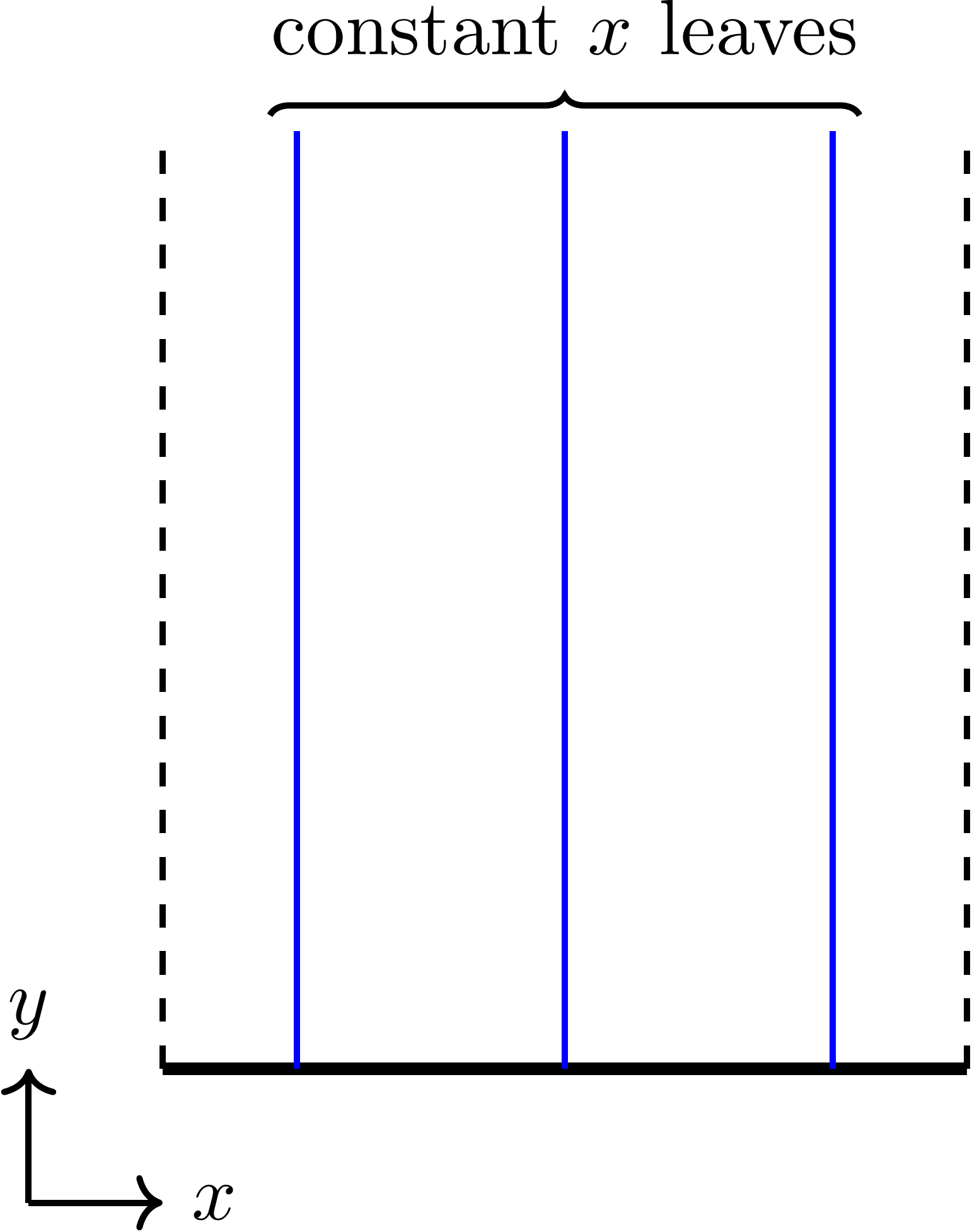}~~~~~~~~~~~
    \includegraphics[scale=0.3]{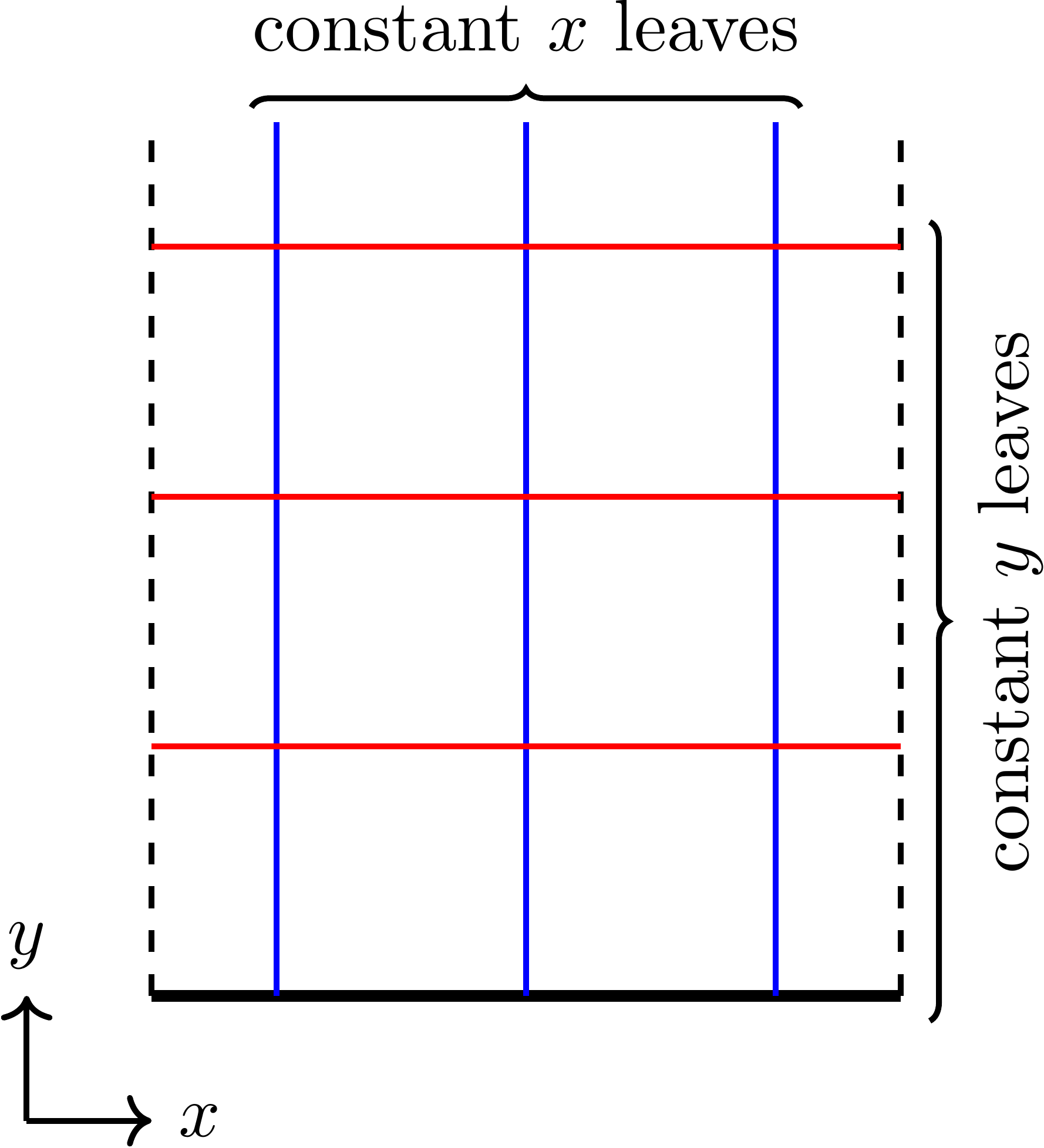}
    \end{center}
    \caption{Two different extensions of the 1+1d boundary (at $y=0$) foliation into the 2+1d bulk. The time direction is not displayed in the figures.}\label{fig:twofoliations}
  \end{figure}

Given a choice of the foliation on the boundary system with subsystem symmetry anomalies, there is typically more than one way to extend the foliation structure into the bulk. 
Consequently, there are generally multiple bulk SSPTs with different bulk foliation structures that can be used to cancel the same boundary anomaly. 
We demonstrate this phenomenon in a  1+1 dimensional system with a $\mathbb{Z}_N$ subsystem symmetry. 
The two different bulk foliation structures are shown in Figure \ref{fig:twofoliations}.
  
\begin{center}
	\emph{Organization}
\end{center}

This paper is organized as follows. In the main text, we will discuss various systems with subsystem global symmetries. We will analyze their 't Hooft anomalies and the corresponding SSPTs in one higher dimension. In parallel, in Appendix \ref{app-A}, we will review analogous systems with ordinary global symmetries, 't Hooft anomalies, and the corresponding SPTs.

Section \ref{sec:U1U1} discusses the anomaly of the $U(1)\times U(1)$ subsystem symmetry and the corresponding 3+1d  SSPT of the 2+1d $\phi$-theory of \cite{paper1}. 
This is to be compared with the mixed anomaly between the momentum $U(1)$ and the winding $U(1)$ symmetry in the ordinary 1+1d compact boson, which we review in Appendix \ref{app:U1U1}. 
In Section \ref{sec:chiral_boson}, we then discuss a $U(1)$ subsystem anomaly in  a chiral version of the scalar field theory in \cite{Gorantla:2020xap}.  
This anomaly is analogous to that of  an ordinary 1+1d chiral boson, discussed in Appendix \ref{app:chiral_boson}.

Section \ref{sec:Xcube} discusses the anomaly and the SSPT of the two $\mathbb{Z}_N$ subsystem global symmetries of the X-cube field theory. 
The discussion is parallel  to that of the $\mathbb{Z}_N\times \mathbb{Z}_N$ one-form global symmetry in the ordinary 2+1d $\mathbb{Z}_N$ gauge theory, the low-energy limit of the toric code. 
We will review this one-form symmetry anomaly in Appendix \ref{app:2+1d_ZN_gauge_theory}.

In Section \ref{sec:2dZN}, we turn to a 1+1d system with a $\mathbb{Z}_N$ subsystem symmetry.   Its 't Hooft anomaly can be canceled by two distinct 2+1d SSPTs with different foliation structures.

Finally, in Section \ref{sec:time} we consider the 2+1d $U(1)$ tensor gauge theory of \cite{paper1} with a $\theta$-angle. 
At $\theta=\pi$, there is a mixed anomaly between a $U(1)$ subsystem symmetry and the time-reversal symmetry. 
This is analogous to the mixed anomaly between the $U(1)$ one-form symmetry and the time-reversal symmetry in the ordinary 1+1d $U(1)$ gauge theory \cite{Gaiotto:2017yup,Komargodski:2017dmc}, which we review in Appendix \ref{app:1+1d_U(1)_gauge_theory}.

\section{Anomalies of $U(1)$ subsystem symmetries}

\subsection{2+1d $U(1)\times U(1)$ subsystem symmetry}\label{sec:U1U1}

It is well-known that the 1+1d  compact boson conformal field theory (CFT), which describes   the gapless phase of the 1+1d XY model, has a $U(1)\times U(1)$ mixed 't Hooft anomaly.  
A mixed 't Hooft anomaly between two global symmetries means that gauging one symmetry breaks the other, and vice versa. The mixed anomaly of the 1+1d compact boson can be canceled by a 2+1d $U(1)\times U(1)$ SPT, whose classical action is given by the mixed Chern-Simons term  (see Appendix \ref{app:anom_compact boson} for a review). 
Here, we demonstrate an analogous  anomaly  for a $U(1)\times U(1)$ subsystem symmetry in the 2+1d $\phi$-theory of \cite{paper1}.  
This anomaly has previously been discussed in  \cite{Gorantla:2021svj,Gorantla:2021bda}.  Below, we review the nature of the anomaly, and present the 3+1d SSPT that cancels it.

\subsubsection{2+1d $\phi$-theory}
The 2+1d $\phi$-theory has a Euclidean Lagrangian
\ie\label{eq:2+1dphi_Lag}
{\cal L}^{(\phi)}_{2+1}={\mu_0\over2} (\partial_\tau\phi)^2 +{1\over 2\mu} (\partial_x\partial_y\phi)^2\,.
\fe
The field $\phi$ is subject to the identification:
\ie
\phi(\tau,x,y)\sim \phi(\tau,x,y) +2\pi n^x(x)+2\pi n^y(y)\,,\quad n^i(x^i)\in \bZ\,.
\fe
Because of this, there exist nontrivial winding configurations of $\phi$ and they are summed over in the path integral (see \cite{paper1} for details).\footnote{An example of winding configurations of $\phi$ on a torus of size $\ell_x,\ell_y$ is
	\ie\label{eq:2+1d_phi_winding}
	\phi = 2\pi\left[\frac{x}{\ell_x}\Theta(y-y_0)+\frac{y}{\ell_y}\Theta(x-x_0)-\frac{xy}{\ell_x\ell_y}\right]~,
	\fe
where $\Theta(x)$ is the Heaviside step function.} This theory describes the gapless phase of the 2+1d XY-plaquette model \cite{PhysRevB.66.054526}.

The 2+1d $\phi$-theory has a $U(1)$ momentum subsystem symmetry that shifts\footnote{Here, by momentum, we mean the conjugate momentum of the field $\phi$ in the target space as opposed to the momentum in coordinate space. Indeed, the temporal current $J_\tau$ of the momentum symmetry is the conjugate momentum of $\phi$.}
\ie\label{U1momentum}
\phi(\tau,x,y)\rightarrow \phi(\tau,x,y) +\alpha_x(x)+\alpha_y(y)~.
\fe
The symmetry is generated by the current
\ie
&J_\tau = i\mu_0 \partial_\tau \phi \,,\qquad J^{xy} = {i\over \mu }\partial^x\partial^y\phi\,,\\
&\partial_\tau J_\tau = \partial_x\partial_y J^{xy}\,.
\fe
The theory also has a $U(1)$ winding subsystem symmetry generated by the current
\ie
&\tilde J_\tau ={1\over 2\pi} \partial_x\partial_y \phi\,,\qquad \tilde J^{xy} = {1\over 2\pi} \partial_\tau\phi\,,\\
&\partial_\tau \tilde J_\tau = \partial_x \partial_y \tilde J^{xy}\,.
\fe
The winding subsystem symmetry does not act on the fundamental field $\phi$, but there are (discontinuous) winding configurations, such as \eqref{eq:2+1d_phi_winding}, carrying nontrivial charge under this symmetry.
This action can be seen explicitly in a dual version of the model, where the winding symmetry shifts the field dual to $\phi$ in a way similar to \eqref{U1momentum}. 
We refer the readers to  \cite{paper1} for details.

The momentum and winding $U(1)$ symmetries can be coupled  to background tensor gauge fields $(A_\tau,A_{xy})$ and $(\tilde A_\tau,\tilde A_{xy})$. The Lagrangian after coupling becomes
\ie\label{LcoupletoA}
\ &{\cal L}^{(\phi)} _{2+1}[A_\tau,A_{xy};\tilde A_\tau,\tilde A_{xy}]
\\
=\ &{\mu_0\over2} (\partial_\tau\phi - A_\tau )^2 +{1\over 2\mu} (\partial_x\partial_y\phi- A_{xy})^2
+{i\over 2\pi} \tilde A_\tau( \partial_x\partial_y \phi  - A_{xy} ) +{i\over 2\pi}\tilde A_{xy} (\partial_\tau\phi -A_\tau)\,.
\fe
It is not invariant under the two $U(1)$ gauge transformations
\ie
&A_\tau \sim A_\tau +\partial_\tau \alpha\,,\qquad A_{xy}\sim A_{xy}+\partial_x\partial_y \alpha\,,\qquad \phi\sim\phi+\alpha\\
&\tilde A_\tau \sim\tilde A_\tau +\partial_\tau\tilde \alpha\,,\qquad \tilde A_{xy}\sim\tilde A_{xy}+\partial_x\partial_y \tilde\alpha\,.
\fe
Rather, it is shifted by
\ie\label{phianomaly}
{\cal L}^{(\phi)}_{2+1}\rightarrow {\cal L} ^{(\phi)}_{2+1}
+{i\over 2\pi}\tilde\alpha ( \partial_\tau  A_{xy} - \partial_x\partial_y A_\tau)\,.
\fe
As discussed in the introduction, we are always free to add 2+1d counterterms involving just the background gauge fields $(A_\tau,A_{xy}), (\tilde A_\tau, \tilde A_{xy})$ to the Lagrangian \eqref{LcoupletoA}.  
However, there is no way to completely remove the anomalous gauge transformation \eqref{phianomaly} by adding these 2+1d local counterterms.  
This signals a mixed 't Hooft anomaly between the $U(1)$ momentum and winding symmetries.

We emphasize that this mixed 't Hooft anomaly is absent in the original 2+1d XY-plaquette lattice model, since the winding subsystem symmetry is only emergent in the low-energy limit. 
On the other hand, it is possible to realize both momentum and winding subsystem symmetry, as well as their mixed 't Hooft anomaly, exactly on a 3-dimensional lattice, where the third direction corresponds to discrete time \cite{Gorantla:2021svj}. Because of the mixed 't Hooft anomaly, the long-distance theory of the latter lattice model is always gapless and is described by the 2+1d $\phi$-theory in the continuum.

\subsubsection{3+1d SSPT}\label{sec:U1U1SSPT}

\begin{figure}
\begin{center}
\includegraphics[scale=.2]{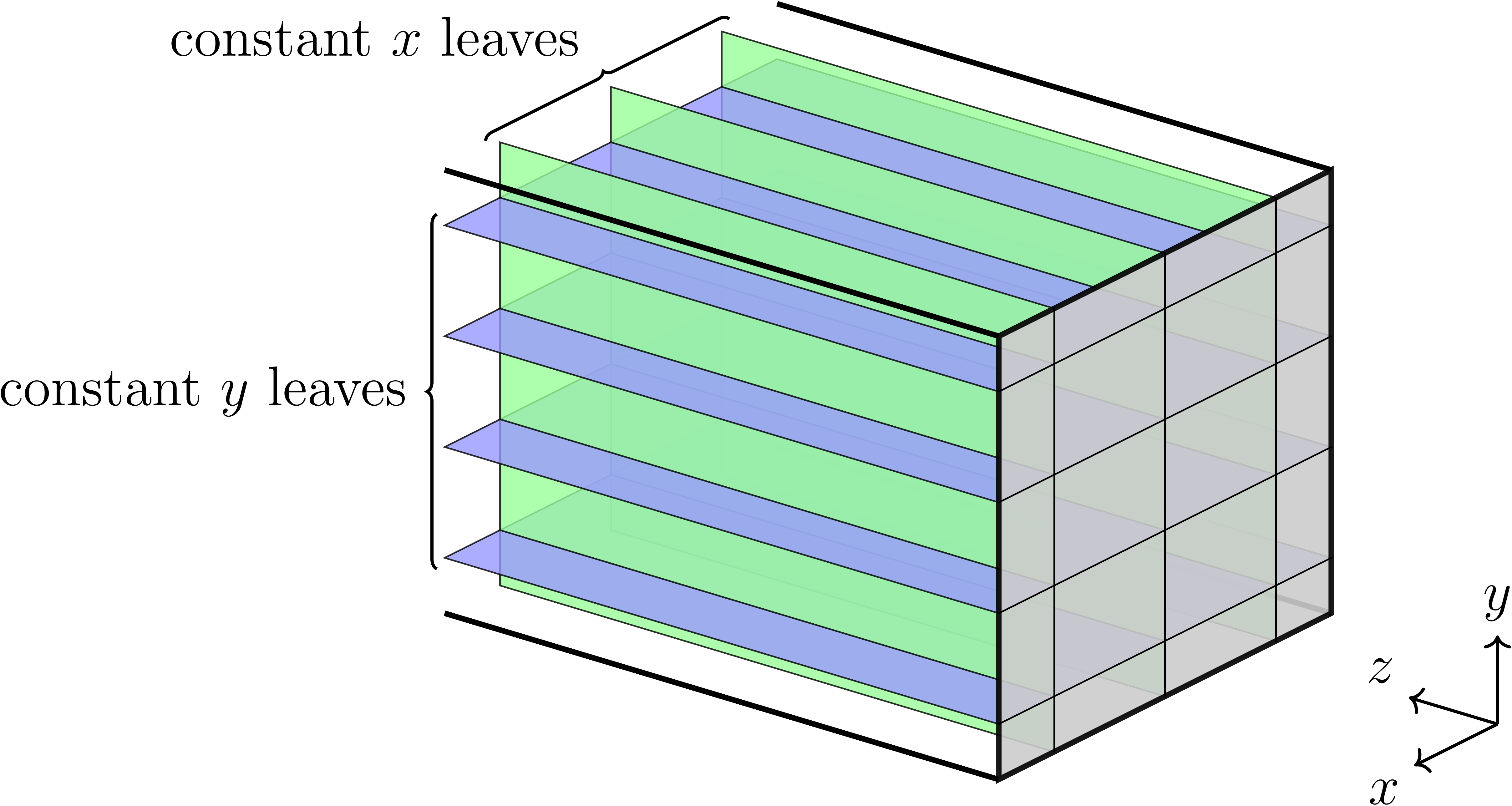}
\end{center}
\caption{The  foliation on the 2+1d boundary (at $z=0$)  is extended to the 3+1d bulk. The time direction is not displayed in the figure.}\label{fig:foliation}
\end{figure}

The mixed anomaly \eqref{phianomaly} can be canceled by coupling the theory to a 3+1d  SSPT.  Denote the radial bulk coordinate by $z$.
The 3+1d geometry will be taken to be $S^1_\tau \times \mathbb{R}_{z\ge 0} \times \Sigma$, with the 2+1d  $\phi$-theory living on the boundary $S^1_\tau \times \Sigma$ at $z=0$.  
Here $\Sigma$ is a 2-manifold with a foliation structure. 
The leaves of the foliation on $\Sigma$ are specified by either the constant $x$ or constant $y$ conditions.\footnote{For example, we can take $\Sigma$ to be a rectangular torus. More generally, $\Sigma$ can be a twisted torus with a choice of the $X$ and $Y$ cycles that wrap finitely many times. See \cite{Rudelius:2020kta} for a related discussion.} 
We extend the foliation structure of $\Sigma$ into the bulk, but we do not introduce additional leaves specified by constant $z$ (see Figure \ref{fig:foliation}). 
For this reason, the bulk SSPT will be called a \textit{2-foliated} SSPT. 

The 3+1d SSPT is protected by a $U(1)\times U(1)$ subsystem symmetry whose conserved charges are supported on the constant $x$ leaves and the constant $y$ leaves at a fixed time. The subsystem symmetry can be coupled to background gauge fields in the bulk with gauge transformations
\ie
&A_\tau \sim A_\tau +\partial_\tau \alpha\,,\qquad A_{xy}\sim A_{xy}+\partial_x\partial_y \alpha\,,\qquad A_z\sim A_z+\partial_z\alpha~,\\
&\tilde A_\tau \sim\tilde A_\tau +\partial_\tau\tilde \alpha\,,\qquad \tilde A_{xy}\sim\tilde A_{xy}+\partial_x\partial_y \tilde\alpha\,,\qquad \tilde A_z \sim \tilde A_z+\partial_z\tilde\alpha\,.
\fe
The components $A_z,\tilde A_z$ are the analogs of the radial components of the bulk gauge fields in ordinary anomaly inflow.

The 3+1d SSPT is described by the classical Euclidean Lagrangian of the background gauge fields:
\ie\label{SSPT_anistropic}
&\mathcal{L}_{3+1}[A_\tau,A_{xy},A_z; \tilde A_\tau,\tilde A_{xy},\tilde A_z] \\
&= {i\over 2\pi} \left[
\tilde A_{xy} (\partial_\tau A_z - \partial_z A_\tau ) +\tilde A_z (\partial_\tau A_{xy}  -\partial_x \partial_y A_\tau)
- \tilde A_\tau (\partial_z A_{xy} - \partial_x \partial_y A_z)
\right]~.
\fe
Under gauge transformations, the Lagrangian is shifted by
\ie
\mathcal{L}_{3+1}\rightarrow \mathcal{L}_{3+1}&+\frac{i}{2\pi}\partial_x[\partial_y \tilde \alpha(\partial_\tau A_z-\partial_z A_\tau)]-\frac{i}{2\pi}\partial_y[ \tilde \alpha\partial_x(\partial_\tau A_z-\partial_z A_\tau)]
\\
&+\frac{i}{2\pi}\partial_z\left[\tilde \alpha (\partial_\tau A_{xy}  -\partial_x \partial_y A_\tau)\right]-\frac{i}{2\pi} \partial_\tau\left[\tilde \alpha (\partial_z A_{xy} - \partial_x \partial_y A_z)\right]\,.
\fe
The bulk action  is invariant up to a boundary term at $z=0$: 
\ie
S_{3+1}=\int d\tau dx dydz \, \mathcal{L}_{3+1}
\rightarrow S_{3+1} -{i\over 2\pi}\int_{z=0} d\tau dxdy \, \tilde \alpha(\partial_\tau A_{xy} -\partial_x\partial_y A_\tau)
\fe
which cancels the anomaly \eqref{phianomaly} of the boundary $\phi$-theory. 

We can also place this SSPT on a 3+1d geometry with only a boundary at $x=0$ or $y=0$. 
We will not discuss the anomaly inflow for those boundaries here.

If we set $\tilde A =A$ in \eqref{SSPT_anistropic}, we will find that the Lagrangian is a total derivative, i.e., there is no Chern-Simons like terms for $(A_\tau, A_{xy}, A_z)$. This implies that the diagonal $U(1)$ subsystem symmetry of the 2+1d $\phi$-theory is anomaly free.

\subsection{3+1d $U(1)$ subsystem symmetry}\label{sec:chiral_boson}

The compact boson CFT in 1+1d has a chiral counterpart, whose anomaly is well-known (see Appendix \ref{app:chiral_boson} for a review).
Here, we show that an analogous theory of  chiral bosons exists in 3+1d, with $U(1)$ subsystem symmetry acting along lines. It can be viewed as a chiral version of the 3+1d $\varphi$-theory of \cite{Gorantla:2020xap} (see also \cite{Yamaguchi:2021qrx}).\footnote{The 2+1d $\phi$-theory \eqref{eq:2+1dphi_Lag} does not have a chiral counterpart, since the naive chiral Lagrangian is a total derivative. 
}
We show that, like its 1+1d cousin, this theory has an 't Hooft anomaly, which can be cancelled by a 4+1d bulk SSPT.

\subsubsection{3+1d chiral $\varphi$-theory}
The 3+1d (non-chiral) $\varphi$-theory of \cite{Gorantla:2020xap} has the Euclidean Lagrangian
\ie
{\cal L}={\mu_0\over2} (\partial_\tau\varphi)^2 +{1\over 2\mu} (\partial_x\partial_y\d_z\varphi)^2\,.
\fe
The field $\varphi$ is subject to the identification:
\ie\label{eq:3+1dvarphi_identification}
\varphi(\tau,x,y,z)\sim \varphi(\tau,x,y,z) +2\pi n^{xy}(x,y)+2\pi n^{yz}(y,z)+2\pi n^{zx}(z,x)\,,~~~~n^{ij}(x^i,x^j)\in \bZ\,.
\fe
Because of this, there exist nontrivial winding configurations for $\varphi$ (see \cite{Gorantla:2020xap} for more details).\footnote{
An example of winding configurations of $\varphi$ on a torus of size $\ell_x,\ell_y,\ell_z$ is
\ie\label{eq:3+1d_varphi_winding}
\varphi = 2\pi&\left[\frac{xyz}{\ell_x\ell_y\ell_z}-\frac{yz}{\ell_y\ell_z}\Theta(x-x_0)-\frac{xz}{\ell_x\ell_z}\Theta(y-y_0)-\frac{xy}{\ell_x\ell_y}\Theta(z-z_0)\right.
\\
&\left.\ \ + \frac{x}{\ell_x}\Theta(y-y_0)\Theta(z-z_0)+\frac{y}{\ell_y}\Theta(x-x_0)\Theta(z-z_0)+\frac{z}{\ell_z}\Theta(x-x_0)\Theta(y-y_0)\right]~.
\fe}

Here we consider a chiral version of this theory with the Euclidean Lagrangian
\ie
{\cal L}^{(\varphi)}_{3+1}=\frac{iN}{4\pi}\d_\tau\varphi\d_x\d_y\d_z\varphi ~,
\fe
where $N\in\mathbb{Z}$. The field $\varphi$ obeys the same identification \eqref{eq:3+1dvarphi_identification} and has the same winding configuration, such as \eqref{eq:3+1d_varphi_winding}, as in the non-chiral theory. The theory has a gauge symmetry
\ie
\varphi(\tau,x,y,z)\sim \varphi(\tau,x,y,z)+g_{xy}(\tau,x,y)+g_{yz}(\tau,y,z)+g_{zx}(\tau,z,x)~.
\fe

The theory also has a momentum subsystem global symmetry that shifts
\ie
\varphi(\tau,x,y,z)\rightarrow \varphi(\tau,x,y,z)+f(x,y,z)~,
\fe
where $f(x,y,z)$ obeys the same global conditions as $\varphi$ does. The symmetry is generated by the current
\ie
&J_\tau=-\frac{N}{2\pi}\d_x\d_y\d_z\varphi~,
\\
&\d_\tau J_\tau=0~.
\fe

We can couple the current $J_\tau$ to a $U(1)$ background tensor gauge field $(A_\tau,A_{xyz})$. This modifies the Lagrangian according to:
\ie \label{Eq:ChiralL}
{\cal L}^{(\varphi)}_{3+1}[A_\tau,A_{xyz}]=\frac{iN}{4\pi}\d_\tau\varphi\d_x\d_y\d_z\varphi-\frac{iN}{2\pi} A_\tau\d_x\d_y\d_z\varphi+\frac{iN}{4\pi} A_\tau A_{xyz}~.
\fe
Note that since the current only has a $J_\tau$ component, the background gauge fields $A_{xyz}$ does not couple to any current. Here we include in the Lagrangian a classical counterterm $\frac{iN}{4\pi} A_\tau A_{xyz}$ for later convenience. This does not affect the 't Hooft anomaly.

The Lagrangian (\ref{Eq:ChiralL})  is not invariant under the $U(1)$ gauge transformation
\ie
\varphi\sim\varphi +\alpha~,\quad (A_\tau,A_{xyz}) \sim (A_\tau+\d_\tau\alpha,A_{xyz}+\d_x\d_y\d_z\alpha)~.
\fe
Rather, it transforms as
\ie\label{eq:anomalyU1}
{\cal L}^{(\varphi)}_{3+1}[A_\tau,A_{xyz}] \to {\cal L}^{(\varphi)}_{3+1}[A_\tau,A_{xyz}]-\frac{iN}{4\pi}\alpha(\d_\tau A_{xyz}-\d_x\d_y\d_z A_\tau)\,.
\fe
It is straightforward to see that no 3+1d local counterterms of $(A_\tau,A_{xyz})$ can be added to cancel the anomalous gauge transformation \eqref{eq:anomalyU1}. 
This signals an 't Hooft anomaly of the $U(1)$ subsystem symmetry.

\subsubsection{4+1d SSPT}

Just as the anomaly of a 1+1d chiral boson can be canceled by a 2+1d $U(1)$ SPT described by a classical Chern-Simons action, we now show that the anomaly \eqref{eq:anomalyU1} can be canceled by coupling the theory to a 4+1d SSPT described by a classical Chern-Simons-like action. Denote the radial bulk coordinate by $w\geq 0$. We will place our chiral $\varphi$-theory at the $w=0$ boundary.

The 4+1d SSPT is protected by a $U(1)$ subsystem symmetry whose conserved charges are supported on either $wx$-, $wy$-, or $wz$-planes.  As in our previous example (see Figure \ref{fig:foliation}), these bulk planes can be viewed as a minimal extension of the boundary foaliation to the bulk, with no additional leaves added parallel to the boundary.  The subsystem symmetry can be coupled to background gauge fields
\ie
(A_\tau,A_{xyz},A_w) \sim (A_\tau+\d_\tau\alpha,A_{xyz}+\d_x\d_y\d_z\alpha,A_w+\d_w\alpha)~.
\fe
The 4+1d SSPT is described by the classical Euclidean Lagrangian
\ie
&\mathcal{L}_{4+1}[A_\tau,A_{xyz},A_w]
\\
&=\frac{iN}{4\pi}
\big[A_\tau(\d_w A_{xyz}-\d_x\d_y\d_z A_w)-A_w(\d_\tau A_{xyz}-\d_x\d_y\d_z A_\tau)+A_{xyz}(\d_\tau A_w-\d_w A_{\tau})\big]~.
\fe
Under the gauge transformation, the Lagrangian is shifted by
\ie
&\frac{iN}{4\pi}
\big[\d_\tau(\alpha\d_w A_{xyz})-\d_w(\alpha\d_\tau A_{xyz})\big]
\\
&+\frac{iN}{4\pi}\big[\d_x(\d_y\d_z\alpha\d_\tau A_w)-\d_y(\d_z\alpha\d_\tau\d_x A_w)+\d_z(\alpha\d_\tau\d_x\d_y A_w)-\d_\tau(\alpha\d_x\d_y\d_z A_w)\big]
\\
&-\frac{iN}{4\pi}\big[\d_x(\d_y\d_z\alpha\d_w A_\tau)-\d_y(\d_z\alpha\d_w\d_x A_\tau)+\d_z(\alpha\d_w\d_x\d_y A_\tau)-\d_w(\alpha\d_x\d_y\d_z A_\tau)\big]
~.
\fe
The bulk action is invariant up to a boundary term at $w=0$:
\ie
S_{4+1}[A_\tau,A_{xyz},A_w]\to S_{4+1}[A_\tau,A_{xyz},A_w]+{iN\over 4\pi}\int_{w=0} d\tau dxdydz \, \alpha(\d_\tau A_{xyz}-\d_x\d_y\d_z A_\tau)~.
\fe
which cancels the anomaly \eqref{eq:anomalyU1} of the boundary chiral $\varphi$-theory.

\section{Anomalies of $\mathbb{Z}_N$ subsystem symmetries}

\subsection{3+1d X-cube field theory}\label{sec:Xcube}

We now consider the continuum field theory \cite{Slagle:2017wrc,paper3} that describes the low-energy physics of the $\mathbb{Z}_N$ X-cube model \cite{Vijay:2016phm}.  
The low-energy field theory has two kinds of $\mathbb{Z}_N$ subsystem global symmetries, one supported  along strips, and the other supported along lines \cite{paper3}. 
The corresponding subsystem global symmetry operators of the continuum field theory descend from the logical operators of the lattice X-cube model.

We will show that there is a mixed 't Hooft anomaly between the two $\mathbb{Z}_N$ subsystem symmetries.  
One manifestation of the anomaly is  that the two subsystem symmetry operators fail to commute when the strips and the lines intersect in space.  
The states in the Hilbert space have to transform under representations of this nontrivial algebra. 
In particular, the sub-extensive ground state degeneracy of the X-cube field theory can  be viewed as a direct consequence of this mixed anomaly \cite{paper3}.

This mixed anomaly is analogous to that between the two $\mathbb{Z}_N$ one-form symmetries of the 2+1d $\mathbb{Z}_N$ gauge theory.  
Both the symmetry operators and the charged objects of the one-form symmetries are the Wilson lines of the 2+1d $\mathbb{Z}_N$ gauge theory. 
The one-form symmetry operators  in the low-energy field theory descend from the string-like logical operators of the microscopic toric code \cite{Kitaev:1997wr}. 
Another manifestation of this anomaly is the nontrivial braiding between the electric and the magnetic Wilson lines. 
We will  review this mixed anomaly in the ordinary 2+1d $\mathbb{Z}_N$ gauge theory in Appendix \ref{app:2+1d_ZN_gauge_theory}. 

Below we will encounter gauge fields transforming in tensor representations of the spatial $S_4$ rotation symmetry. These can be described with spatial indices $(i,j,k)$, which we will always take to be cyclically-ordered and, in particular, non-equal, i.e., $(i,j,k)=(x,y,z)$, $(y,z,x)$, or $(z,x,y)$.  We will use $[ij]$ to denote anti-symmetrization over the indices $i$ and $j$, and $(ij)$ to indicate symmetrization thereof.
 The tensors we will encounter include the symmetric off-diagonal tensor $T^{ij}=T^{ji}$ with three component $T^{xy},T^{yz},T^{zx}$; the partially anti-symmetric tensor $T^{[ij]k}$, with three components $T^{[xy]z},T^{[yz]x},T^{[zx]y}$ and a common gauge symmetry $T^{[ij]k}\sim T^{[ij]k}+c$; the partially symmetric tensor $T^{i(jk)}$  with three components that obey $T^{x(yz)}+T^{y(zx)}+T^{z(xy)}=0$; and $T_{[ij]k}$ with three components that obey $T_{[xy]z}+T_{[yz]x}+T_{[zx]y}=0$. Note that we distinguish upper and lower indices. Fields with indices $T^{[ij]k}$ and $T^{k(ij)}$ are related by $T^{k(ij)}\equiv T^{[ki]j} - T^{[jk]i}$.   
 For more details on this notation and its connection to representations of spatial $S_4$ rotation group, see \cite{paper2,paper3,Gorantla:2021svj}.

The $\mathbb{Z}_N$ X-cube model can be described by a low-energy continuum field theory using two sets of $U(1)$ tensor gauge fields
\ie \label{Eq:XcubeGaugeT}
&
(A_\tau,A_{ij}) \sim (A_\tau+\partial_\tau\alpha, A_{ij}+\d_i\d_j\alpha)~,
\\
&
(\hat A_\tau^{i(jk)},\hat A^{ij}) \sim (\hat A_\tau^{i(jk)}+\d_\tau\hat \alpha^{i(jk)},\hat A^{ij}+\d_k\hat\alpha^{k(ij)})~.
\fe
The Euclidean Lagrangian of the low-energy continuum field theory is \cite{Slagle:2017wrc,paper3}
\ie\label{eq:LagXcube}
\mathcal{L}^\text{XC}_{3+1}=\frac{iN}{2\pi}\big[A_{ij}(\d_\tau\hat A^{ij}-\d_k\hat A_\tau^{k(ij)})+A_\tau\partial_i\partial_j \hat A^{ij}\big]~.
\fe

The simplest gauge-invariant  operator that can be constructed from $A$ is
\ie
W(x,y,z)=\exp\left[i\oint d\tau A_\tau\right]~,
\fe
which is the worldline of an immobile fracton. 
In addition, we have 
\ie \label{Eq:wstrip}
W(z_1,z_2,\mathcal{C})=\exp\left[i\int_{z_1}^{z_2}dz\oint_\mathcal{C}(\d_zA_\tau\,d\tau+A_{zx}\,dx+ A_{yz}\,dy)\right]~,
\fe
where $\mathcal{C}$ is a curve in $(\tau, x,y)$ representing the world strip of a dipole of fractons separated in the $z$-direction, which are mobile in the $(x,y)$-plane. 

The gauge-invariant operators constructed from $\hat A$ have the form
\ie \label{Eq:What}
\hat{W}^z(x,y,\hat{\mathcal{C}})=\exp\left[i\oint_{\hat{\mathcal{C}}} (\hat A_\tau^{z(xy)}d\tau+ \hat A^{xy}\,dz)\right]~,
\fe
where $\hat{\mathcal{C}}$ is a curve in the $(\tau, z)$-plane, which describes the world-line of a $z$-lineon, which can move only along the $z$-direction. The analogous operators $\hat{W}^x(y,z,\hat{\mathcal{C}})$ and $\hat{W}^y(z,x,\hat{\mathcal{C}})$ represent the motion of $x$- and $y$-lineons.  

\subsubsection{Global symmetries and anomaly}

We now discuss the two $\mathbb{Z}_N$ subsystem global symmetries of the continuum field theory for the X-cube model.

\begin{center}
	\emph{$\mathbb{Z}_N$ subsystem global symmetries}
\end{center}
The theory has a $\mathbb{Z}_N$ subsystem global symmetry   generated by the $\hat W$ operators in Eq. (\ref{Eq:What}), with   $\hat {\cal C}$ chosen to be a straight line at a fixed time.  Due to the commutation relations implied by (\ref{eq:LagXcube}), such an operator shifts $(A_\tau,A_{ij})$ by a flat $\mathbb{Z}_N$ tensor gauge field. 
The corresponding gauge-invariant charged operators  are the $W$ operators.

We can couple the symmetry to the following background tensor gauge fields\footnote{The $U(1)$ and the $\mathbb{Z}_N$ tensor gauge theories of $(C_{\tau ij},C_{[ij]k})$ were studied in \cite{Gorantla:2020xap}.}
\ie \label{Eq:Cgauge}
(C_{\tau ij},C_{[ij]k}) \sim (C_{\tau ij}+\partial_\tau\lambda_{ij}-\partial_i\partial_j\lambda_\tau,C_{[ij]k}-\d_i\lambda_{jk}+\d_j\lambda_{ik})~.
\fe
The gauge parameters $(\lambda_\tau,\lambda_{ij})$ have their own gauge symmetry
\ie
(\lambda_\tau,\lambda_{ij}) \sim (\lambda_\tau+\d_\tau\gamma,\lambda_{ij}+\d_i\d_j\gamma)~.
\fe
These background gauge fields are $U(1)$ gauge fields. We can restrict them to $\mathbb{Z}_N$ gauge fields by coupling them to the dynamical fields $\hat \phi^{[ij]k}$ :
\ie
\mathcal{L}_C=\frac{iN}{2\pi}\hat \phi^{[ij]k}(\d_\tau C_{[ij]k}+\d_iC_{\tau jk}-\d_jC_{\tau ik})~.
\fe

The theory  has another $\mathbb{Z}_N$ subsystem global symmetry   generated by the $W$ operator (\ref{Eq:wstrip}) with  $\cal C$ chosen to be at a fixed time.  This operator shifts $(\hat A_\tau^{k(ij)},\hat A^{ij})$ by a flat $\mathbb{Z}_N$ tensor gauge field. 
The corresponding gauge-invariant charged operators   are the $\hat W$ operators.

We can couple this second symmetry to the following background tensor gauge fields\footnote{The $U(1)$ and the $\mathbb{Z}_N$ tensor gauge theories of $(\hat C_\tau^{ij},\hat C)$ were studied in \cite{Gorantla:2020xap}.}
\ie \label{Eq:Cgauge2}
(\hat C_\tau^{ij},\hat C) \sim (\hat C_\tau^{ij}+\partial_\tau\hat \lambda^{ij}-\partial_k\hat\lambda_\tau^{k(ij)},\hat C+\d_i\d_j\hat\lambda^{ij})~,
\fe
where the gauge parameters $(\hat \lambda_\tau^{k(ij)},\hat\lambda^{ij})$ have their own gauge symmetry
\ie
(\hat \lambda_\tau^{k(ij)},\hat\lambda^{ij}) \sim (\hat \lambda_\tau^{k(ij)}+\d_\tau\hat\gamma^{k(ij)},\hat\lambda^{ij}+\d_k\hat\gamma^{k(ij)})~.
\fe
Again, these background gauge fields are $U(1)$ gauge fields. To restrict them to $\mathbb{Z}_N$ gauge fields, we couple them to a dynamical scalar $\phi$:
\ie
\mathcal{L}_{\hat C}=\frac{iN}{2\pi}\phi(\d_\tau \hat C-\d_i\d_j\hat C_\tau^{ij})~.
\fe

\begin{center}
	\emph{Mixed 't Hooft anomaly}
\end{center}

Having introduced the background gauge fields for the two $\mathbb{Z}_N$ subsystem symmetries, we now discuss how they are coupled to the X-cube model at low energy.
This coupling is described by the Lagrangian 
\ie
&\,\mathcal{L}_{3+1}^\text{XC}[C,\hat C]
\\
=&\,\mathcal{L}_C+\mathcal{L}_{\hat C}+
\frac{iN}{2\pi}\left[A_{ij}(\d_\tau\hat A^{ij}-\d_k\hat A_\tau^{k(ij)}-\hat C_\tau^{ij})+A_\tau(\partial_i\partial_j \hat A^{ij}-\hat C) +\hat A^{ij}C_{\tau ij}+\hat A_\tau^{[ij]k}C_{[ij]k}\right] 
~.
\fe

To ensure that this Lagrangian is invariant under the dynamical gauge symmetry of the X-cube model, we require that under the gauge transformations (\ref{Eq:XcubeGaugeT}),  
the dynamical scalar fields transform according to:
\ie
&\phi \sim \phi-\alpha~,
\\
&\hat\phi^{[ij]k} \sim \hat\phi^{[ij]k}+\hat\alpha^{[ij]k}~,
\fe
where $\hat \alpha^{[ij]k}$ obeys $\hat \alpha^{k(ij)}=\hat \alpha^{[ki]j}-\hat \alpha^{[jk]i}$ and  has a gauge symmetry $\hat \alpha^{[ij]k}\sim \hat \alpha^{[ij]k}+c$.
Additionally, under the gauge transformations (\ref{Eq:Cgauge}) and (\ref{Eq:Cgauge2}) of the background gauge fields, the dynamical fields of the original X-cube model transform according to:
\ie
&(A_\tau,A_{ij}) \sim (A_\tau+\lambda_\tau,A_{ij}+\lambda_{ij})~,
\\
&(\hat A_\tau^{k(ij)},\hat A^{ij}) \sim (\hat A_\tau^{k(ij)}+\hat \lambda_\tau^{k(ij)},\hat A^{ij}+\hat\lambda^{ij})~.
\fe
Thus we see that under these background gauge transformations, the Lagrangian is not invariant; rather, it is shifted by
\ie\label{eq:anomalyXcube}
\frac{iN}{2\pi}\left[-\lambda_{ij}\hat C_\tau^{ij}-\lambda_\tau\hat C+ \hat \lambda^{ij}\left(C_{\tau ij}+\partial_\tau\lambda_{ij}-\partial_i\partial_j\lambda_\tau\right)+\hat \lambda_\tau^{[ij]k}\left(C_{[ij]k}-\d_i\lambda_{jk}+\d_j\lambda_{ik}\right)\right]~.
\fe
This signals a mixed 't Hooft anomaly between the two $\mathbb{Z}_N$ subsystem global symmetries.

\subsubsection{4+1d SSPT}
The mixed 't Hooft anomaly described by Eq. (\ref{eq:anomalyXcube}) can be canceled by coupling the system to a 4+1d SSPT protected by two $\mathbb{Z}_N$ subsystem symmetries, as we now describe. We denote the radial bulk coordinate by $w\ge0$, and place our X-cube field theory at the $w=0$ boundary.  

Our bulk theory has two sets of background tensor gauge fields, which couple to the two subsystem $\mathbb{Z}_N$ symmetries.   The first set is comprised of the fields
\ie
&(C_{\tau ij},C_{[\tau w]},C_{[ij]k},C_{wij})
\\
&\sim (C_{\tau ij}+\partial_\tau\lambda_{ij}-\partial_i\partial_j\lambda_\tau,C_{[\tau w]}+\d_\tau\lambda_w-\d_w\lambda_\tau,C_{[ij]k}-\d_i\lambda_{jk}+\d_j\lambda_{ik},C_{wij}+\partial_w\lambda_{ij}-\partial_i\partial_j\lambda_w)~,
\fe
As in our boundary theory, the gauge parameters $(\lambda_\tau,\lambda_{ij},\lambda_w)$ have their own gauge symmetry
\ie
(\lambda_\tau,\lambda_{ij},\lambda_w) \sim (\lambda_\tau+\d_\tau\gamma,\lambda_{ij}+\d_i\d_j\gamma,\lambda_w+\d_w\gamma)~.
\fe
Since these are $U(1)$ gauge fields, we couple them to a dynamical gauge field
\ie
(\hat B_\tau^{k(ij)},\hat B^{ij},\hat B_w^{k(ij)}) \sim (\hat  B_\tau^{k(ij)}+\d_\tau\hat\beta^{k(ij)},\hat B^{ij}+\d_k\hat\beta^{k(ij)},\hat B_w^{k(ij)}+\d_w\hat\beta^{k(ij)})~,
\fe
with the Euclidean Lagrangian
\ie
\mathcal{L}_{4+1}^{(C)}=\frac{iN}{2\pi}&\left[\hat B_\tau^{[ij]k}(\d_w C_{[ij]k}+\d_iC_{wjk}-\d_jC_{wik})\right.
\\
&\ \left.-\hat B_w^{[ij]k}(\d_\tau C_{[ij]k}+\d_iC_{\tau jk}-\d_jC_{\tau ik})-\hat B^{ij}(\d_i\d_j C_{[\tau w]}+\d_\tau C_{wij}-\d_w C_{\tau ij})\right]~.
\fe

The second $\mathbb{Z}_N$ symmetry couples to the following tensor gauge fields:
\ie
&(\hat C_\tau^{ij},\hat C_{[\tau w]}^{k(ij)},\hat C,\hat C_w^{ij})
\\
&\sim  (\hat C_\tau^{ij}+\partial_\tau\hat \lambda^{ij}-\partial_k\hat\lambda_\tau^{k(ij)},\hat C_{[\tau w]}^{k(ij)}+\d_\tau\hat \lambda_w^{k(ij)}-\d_w\hat \lambda_\tau^{k(ij)},\hat C+\d_i\d_j\hat\lambda^{ij},\hat C_w^{ij}+\partial_w\hat \lambda^{ij}-\partial_k\hat\lambda_w^{k(ij)})~.
\fe
where the gauge parameters $(\hat \lambda_\tau^{k(ij)},\hat\lambda^{ij},\hat\lambda_w^{k(ij)})$ have their own gauge symmetry
\ie
(\hat \lambda_\tau^{k(ij)},\hat\lambda^{ij},\hat\lambda_w^{k(ij)}) \sim (\hat \lambda_\tau^{k(ij)}+\d_\tau\hat\gamma^{k(ij)},\hat\lambda^{ij}+\d_k\hat\gamma^{k(ij)},\hat\lambda_w^{k(ij)}+\d_w\hat\gamma^{k(ij)})~.
\fe
To restrict the symmetry to $\mathbb{Z}_N$, we introduce the dynamical gauge fields
\ie
(B_\tau ,B_{ij},B_w) \sim (B_\tau +\d_\tau \beta,B_{ij}+\d_i\d_j\beta,B_w+\d_w\beta)~.
\fe
which couple to $\hat{C}$ via the Euclidean Lagrangian
\ie \label{Eq:LagChat}
\mathcal{L}^{(\hat C)}_{4+1}=\frac{iN}{2\pi}&\left[B_\tau (\d_w\hat C-\d_i\d_j\hat C^{ij}_w)-B_w(\d_\tau \hat C-\d_i\d_j\hat C^{ij}_\tau )-B_{ij}(\d_k\hat C_{[\tau w]}^{k(ij)}+\d_\tau \hat C_w^{ij}-\d_w\hat C_\tau ^{ij})\right]~.
\fe

Our 4+1d SSPT is described by coupling these two subsystem-symmetric theories, via the classical Euclidean Lagrangian
\ie
\mathcal{L}_{4+1d}=\mathcal{L}_{4+1}^{(C)}+\mathcal{L}_{4+1}^{(\hat C)}+\frac{iN}{2\pi}\left(C_{\tau ij}\hat C_w^{ij}-C_{wij}\hat C_\tau ^{ij}+C_{[\tau w]}\hat C-C_{[ij]k}\hat C_{[\tau w]}^{[ij]k}\right)
\fe
For the coupling between the $C$ and $\hat C$ gauge fields to be gauge-invariant, the dynamical gauge fields $B$ and $\hat B$ also have to transform under the background gauge symmetries of $(\lambda_\tau ,\lambda_{ij},\lambda_w)$ and $(\hat \lambda_\tau ^{k(ij)},\hat\lambda^{ij},\hat\lambda_w^{k(ij)})$ as follows:
\ie\label{BhatB-bkggaugesym}
&(B_\tau ,B_{ij},B_w) \sim (B_\tau -\lambda_\tau ,B_{ij}-\lambda_{ij},B_w-\lambda_w)~,
\\
&(\hat B_\tau ^{k(ij)},\hat B^{ij},\hat B_w^{k(ij)}) \sim (\hat B_\tau^{k(ij)}+\hat \lambda_\tau ^{k(ij)},\hat B^{ij}+\hat \lambda^{ij},\hat B_w^{k(ij)}+\hat \lambda_w^{k(ij)})~.
\fe
It is straightforward to check that the resulting action is invariant under the background gauge transformations up to a boundary term at $w=0 $, which cancels the anomaly \eqref{eq:anomalyXcube} of the boundary X-cube field theory \eqref{eq:LagXcube}.

\subsection{1+1d system with a $\mathbb{Z}_N$ subsystem symmetry}\label{sec:2dZN}

We now consider a 1+1d system with a $\mathbb{Z}_N$ subsystem symmetry.  As we show in Section \ref{sec:gaugechiral}, this system is related to the 1+1d chiral boson by gauging a $\mathbb{Z}$ subsystem symmetry.   

Consider the Euclidean action
\ie\label{eq:action_subsys}
S = \frac{iN}{4\pi}\oint dx\left[\int_{\tau_*}^{\tau_* +\ell_\tau} d\tau \, \d_\tau\phi(\tau, x)\d_x\phi(\tau,x) + 2\pi \d_x\phi(\tau_*, x) w_\tau(x)\right]~.
\fe
The field $\phi$ is subject to the identification
\ie\label{eq:period_subsys}
\phi(\tau, x) \sim \phi(\tau, x)+2\pi m(x)~, \qquad m(x) \in \mathbb Z ~.
\fe
It can wind in the Euclidean time and the spatial directions:
\ie\label{eq:winding_subsys}
&\phi(\tau+\ell_\tau,x) = \phi(\tau,x)+2\pi w_\tau(x)~, \qquad &&w_\tau(x) \in \mathbb Z ~,\\
&\phi(\tau,x+\ell_x) = \phi(\tau,x)+2\pi w_x~,\qquad &&w_x \in \mathbb Z ~.
\fe
Here $w_\tau(x)$ is a single-valued, integer function. Indeed, under the identification \eqref{eq:period_subsys}, the action is shifted by
\ie\label{check}
S \rightarrow S+ 2\pi i N \oint dx\, w_\tau(x)\d_xm(x) ~,
\fe
which is an integer multiple of $2\pi i$. Therefore the partition function is invariant under this subsystem symmetry.

The action \eqref{eq:action_subsys} is very similar to that of an ordinary chiral boson \eqref{app-chiralboson-Lag}, but they differ in the second term.  
We now explain the importance of this term. 
Because of the position-dependent winding of $\phi$ in the $\tau$-direction, the first term in the action \eqref{eq:action_subsys} is not well-defined on its own. 
This is precisely  fixed by adding the second correction term. With the correction term, the full action \eqref{eq:action_subsys} is independent of the choice of the reference time $\tau_*$, and it is invariant modulo $2\pi i$ under the identification \eqref{eq:period_subsys} (see \eqref{check}). This correction term is similar to the correction term in the action of the quantum mechanics of $N$ degenerate ground states reviewed in Appendix \ref{app:1+1d_pqdot}. 

In addition to the subsystem symmetry, the theory (\ref{eq:action_subsys}) has a gauge symmetry
\ie\label{eq:1+1d_ZN2_gauge_sym_lattice}
\phi(\tau,x)\sim\phi(\tau,x)+g(\tau)~,
\fe
where $g(\tau)$ can wind in time, $g(\tau+\ell_\tau)-g(\tau)\in 2\pi\mathbb{Z}$.

\subsubsection{Global symmetry and anomaly}

The action (\ref{eq:action_subsys}) is invariant under the $\mathbb{Z}_N$ subsystem global symmetry: 
\ie\label{eq:1+1d_ZN2_global_sym}
\phi(\tau,x)\rightarrow \phi(\tau, x) + \frac{2\pi m(x)}{N}~,\qquad m(x)\in\mathbb{Z}~.
\fe
Note that the symmetry with constant $m(x)=m$ is not a global symmetry, but part of the gauge symmetry  \eqref{eq:1+1d_ZN2_gauge_sym_lattice}. The subsystem global symmetry is generated by the gauge invariant operator
\ie
U(x_1, x_2)=\exp\left[i\phi(x_2)-i\phi(x_1)\right]~,
\fe
with $x_1<x_2$. It shifts $\phi(\tau,x)\rightarrow \phi(\tau,x) +\frac{2\pi}{N}$ only within an interval, $x\in(x_1,x_2)$.

Using the commutation relation,
\ie
\left[\phi(x_1),\d_x \phi(x_2)\right]= \frac{2\pi i}{N}\delta(x_1-x_2)~,
\fe
we find that the symmetry operators obey a nontrivial commutation relation
\ie\label{eq:1+1d_ZN2_commute}
U(x_1,x_2)U(x_3,x_4)=
\begin{cases}
	e^{-2\pi i/N}U(x_3,x_4)U(x_1,x_2) \quad &x_1<x_3<x_2<x_4
	\\
	e^{2\pi i/N}U(x_3,x_4)U(x_1,x_2)\quad & x_3<x_1<x_4<x_2
	\\
	U(x_3,x_4)U(x_1,x_2) \quad &\text{otherwise}
\end{cases}~.
\fe
This signals an 't Hooft anomaly of the $\mathbb{Z}_N$ subsystem symmetry. 
In other words, the   $\mathbb{Z}_N$ subsystem symmetry acts projectively on the Hilbert space.

The 't Hooft anomaly can also be detected by coupling the system to the background gauge field $A_{\tau}$ for the $\mathbb{Z}_N$ subsystem symmetry. The action after coupling becomes
\ie
S=\frac{iN}{4\pi}\oint d\tau dx \left( \d_\tau\phi\d_x \phi -2A_{\tau}\d_x \phi\right)~.
\fe
Here we omit the correction term, and restrict the holonomies of the $U(1)$ background gauge fields $A_\tau$ to be $\mathbb{Z}_N$-valued.
The background gauge symmetry is
\ie
\phi\sim \phi+\alpha~,\quad A_\tau\sim A_\tau +\d_\tau\alpha~.
\fe
It shifts the action by
\ie\label{eq:1+1d_ZN2_anom}
S\rightarrow S-\frac{iN}{4\pi}\oint d\tau dx\,(\d_\tau \alpha\d_x\alpha+2A_\tau \d_x\alpha)~.
\fe
which signals an 't Hooft anomaly of the  $\mathbb{Z}_N$ subsystem symmetry.

\subsubsection{2+1d SSPT}

We now construct the 2+1d SSPTs that cancel the anomaly via anomaly inflow. We will show that there are two such SSPTs, and they differ in their global symmetry and foliation. Denote the radial bulk coordinate by $y\ge0$. We place our 1+1d system on the $y=0$ boundary.

\begin{center}
	\emph{1-foliated SSPT}
\end{center}

First, we consider a 2+1d SSPT protected by a 1-foliated $\mathbb{Z}_N$ subsystem global symmetry. The subsystem symmetry is generated by distinct symmetry line operators, that extend in the $y$ direction, at different $x$. This is illustrated on the left in Figure \ref{fig:twofoliations}. Hence we refer to it as a 1-foliated symmetry, and the corresponding SSPT as 1-foliated SSPT.

We can couple the subsystem symmetry to a background gauge field $(A_{\tau}, A_{y})$. The SSPT that we are interested in is described by the Euclidean Lagrangian
\ie
\mathcal{L}_{2+1}=\frac{iN}{2\pi}\Phi (\d_\tau A_{y}-\d_y A_{\tau})-
\frac{iN}{2\pi}A_{\tau} \d_x A_{y} ~,
\fe
where $\Phi$ is a dynamical field of mass dimension +1 that constrains the classical $U(1)$ gauge field $(A_{\tau}, A_{y})$ to a $\mathbb{Z}_N$ gauge field. The background gauge transformation, 
\ie
\Phi\sim \Phi + \d_x\alpha~,\quad (A_\tau,A_y)\sim (A_\tau+\d_\tau\alpha, A_y+\d_y\alpha)~,
\fe
shifts the Lagrangian by total derivatives
\ie\label{1_foliated_bulk}
-\frac{iN}{4\pi}\left[\d_x(\d_\tau\alpha\d_y\alpha+2 A_y\d_\tau\alpha)+\d_y(\d_\tau\alpha\d_x\alpha+2A_\tau\d_x\alpha)-\d_\tau(\d_x\alpha\d_y\alpha+2A_y\d_x\alpha )\right]
\fe
It leads to an anomaly inflow to the $y=0$ boundary, which cancels the anomaly \eqref{eq:1+1d_ZN2_anom} of the boundary 1+1d system.

\begin{center}
	\emph{2-foliated SSPT}
\end{center}

Second, we show that the anomaly can also be canceled by a 2+1d SSPT protected by a 2-foliated $\mathbb{Z}_N$ subsystem symmetry. The subsystem symmetry is generated by symmetry line operators that extend in either the $x$ or the $y$ direction. This is illustrated on the right in Figure \ref{fig:twofoliations}. These symmetry lines are all distinct operators except that the products of symmetry lines that extend in the $x$ and $y$ directions are the same.

We can couple the subsystem symmetry to a $\mathbb{Z}_N$ background tensor gauge field $(A_{\tau}, A_{xy})$. The SSPT that we are interested in is described by the classical Lagrangian
\ie
\mathcal{L}_{2+1}=\frac{iN}{2\pi}\Phi^{xy} (\d_\tau A_{xy}-\d_x\d_y A_\tau)-\frac{iN}{2\pi}A_\tau A_{xy} ~,
\fe
where $\Phi^{xy}$ is a dimensionless dynamical field that constrains the classical $U(1)$ gauge field $(A_\tau,A_{xy})$ to a $\mathbb{Z}_N$ gauge field. The gauge symmetry is
\ie
A_\tau \sim A_\tau+\d_\tau\alpha~,\qquad A_{xy} \sim A_{xy}+\d_x\d_y\alpha~,\qquad \Phi^{xy} \sim \Phi^{xy}-\alpha~.
\fe
The Lagrangian is invariant under the gauge symmetry up to total derivatives
\ie\label{2_foliated_bulk}
-\frac{iN}{4\pi}\left[\partial_x(\partial_\tau\alpha\partial_y\alpha-2\alpha\partial_yA_\tau)+\partial_y(\partial_\tau\alpha\partial_x\alpha+2A_\tau\partial_x\alpha)-\partial_\tau(\partial_x\alpha\partial_y\alpha-2\alpha A_{xy})\right]~.
\fe
These terms signal an anomaly inflow to the $y=0$ boundary, which cancels the anomaly \eqref{eq:1+1d_ZN2_anom} of the boundary 1+1d system.

\subsubsection{Relation to the ordinary chiral boson}\label{sec:gaugechiral}
We now show that starting from the 1+1d ordinary chiral boson theory (reviewed in Appendix \ref{app:chiral_boson}), we can gauge a $\mathbb{Z}$ subsystem symmetry to obtain the 1+1d system \eqref{eq:action_subsys}.  
We will see that the correction term in \eqref{eq:action_subsys} arises naturally from this gauging.

In the absence of the velocity term  $(\partial_x \varphi)^2$,  the ordinary chiral boson theory \eqref{app-chiralboson-Lag} of $\varphi$ has a  $\mathbb{Z}$ subsystem symmetry that acts as
\ie
\varphi(\tau,x)\rightarrow \varphi(\tau,x)+2\pi w_\tau(x), ~~~w_\tau(x)\in\mathbb{Z}\, .
\fe
Note that the conjugate momentum of $\varphi$ is $\pi=\frac{N}{2\pi}\partial_x\varphi$. Gauging the subsystem symmetry amounts to inserting the symmetry operators
\ie
\exp\left[2\pi i\oint dx\,w_\tau(x)\pi(\tau_*,x)\right]=\exp\left[iN\oint dx\,w_\tau(x) \d_x\varphi(\tau_*,x)\right]~,
\fe
at a given time $\tau_*$, and then summing over all possible $w_\tau(x)$.\footnote{Note that the symmetry operator with constant $w_\tau(x)=n$ is a trivial operator since it shifts $\varphi \rightarrow \varphi +2\pi n$.} Inserting these symmetry operators is equivalent to modifying the action to
\ie
S=\frac{iN}{4\pi}\oint dx\left[\int_{\tau_*}^{\tau_* +\ell_\tau}d\tau \, \d_\tau\varphi(\tau,x)\d_x\varphi(\tau,x)+4\pi w_\tau(x)\d_x\varphi(\tau_*,x)\right]~,
\fe
where we sum over the winding configurations \eqref{app-chiralboson-boundary_condition} of $\varphi$ in the path integral. 

Next, we define a new field $\phi$ as
\ie
\phi(\tau,x) = \varphi(\tau,x)+ 2\pi w_\tau(x)\Theta(\tau-\tau_*)\,,
\fe
and rewrite the path integral using this field. 
In contrast to the old field $\varphi$,  the new field $\phi$ can have nontrivial, $x$-dependent winding in the Euclidean time direction. 
The action then becomes \eqref{eq:action_subsys}, and in the path integral, we now sum over the winding configurations \eqref{eq:winding_subsys} of $\phi$.

\section{Anomalies of time-reversal and subsystem symmetries}\label{sec:time}

For our final example, we consider a time-reversal invariant $U(1)$ tensor gauge theory with a $\theta$-angle $\theta=\pi$.  
We show that this system exhibits is a mixed 't Hooft anomaly between time-reversal symmetry and a subsystem symmetry. 
This anomaly is analogous to the mixed anomaly between the electric one-form symmetry and the time-reversal symmetry in the ordinary 1+1d $U(1)$ gauge theory at $\theta=\pi$ \cite{Gaiotto:2017yup,Komargodski:2017dmc}. See Appendix \ref{app:1+1d_U(1)_gauge_theory} for a review.  
  
Importantly, the tensor gauge fields $(A_\tau,A_{xy})$ below are dynamical gauge fields, rather than background gauge fields for a response action. 
The subsystem symmetry in question is analogous to a  one-form symmetry, whose charged objects are  extended Wilson strips rather than points.

\subsection{2+1d $U(1)$ tensor gauge theory}

Consider a 2+1d $U(1)$ tensor gauge theory of a symmetric tensor gauge field
\ie
(A_\tau,A_{xy}) \sim (A_\tau+\d_\tau\alpha,A_{xy}+\d_x\d_y\alpha)~,
\fe
with a Euclidean Lagrangian \cite{paper1}
\ie \label{Eq:Tlagg1}
\mathcal{L}^{(A)}_{2+1}=\frac{1}{g_e^2}E_{xy}^2+\frac{i\theta}{2\pi}E_{xy}~.
\fe
Here $E_{xy}=\d_\tau A_{xy}-\d_x\d_y A_\tau $ is the gauge-invariant electric field. 
The $\theta$-parameter has a $2\pi$ periodicity, i.e., $\theta\sim \theta+2\pi$, since the electric flux is quantized as follows:
\ie
\oint d\tau dx dy\, E_{xy}\in 2\pi\mathbb{Z}~.
\fe
To see this quantization explicitly, observe that the gauge field configuration 
\ie\label{eq:2+1d_gauge_flux}
A_\tau=0~, \quad A_{xy} = \frac{2\pi \tau}{\ell_\tau}\left[\frac{1}{\ell_x}\delta(y-y_0)+\frac{1}{\ell_y}\delta(x-x_0)-\frac{1}{\ell_x\ell_y}\right]~
\fe
carries nontrivial electric flux.
This is a valid configuration since $A_{xy}(\tau+\ell_\tau,x,y)-A_{xy}(\tau,x,y)=\d_x\d_y g$ with the transition function $g$ given by  \eqref{eq:2+1d_phi_winding}.  See \cite{paper1} for a more detailed discussion of the fluxes in this $U(1)$ tensor gauge theory. 

 Because $\theta$ is periodic, the theory at $\theta=0,\pi$ has a time-reversal symmetry
\ie
(\tau, x,y)\rightarrow (-\tau,x,y)~,\quad (A_\tau,A_{xy})\rightarrow(-A_\tau,A_{xy})~.
\fe

If the gauge field is classical, the $\theta$-term at $\theta=\pi$ can be viewed as a response action for a 2+1d SSPT with gapless corner modes protected by time-reversal and $U(1)$ subsystem symmetry \cite{You:2019bvu}.   
Here, we instead consider the situation where the tensor gauge field is dynamical.

In this case, at all values of $\theta$, the theory has a $U(1)$ electric tensor symmetry that shifts $(A_\tau,A_{xy})$ by a flat tensor gauge field. The symmetry is generated by a current
\ie
&J_\tau ^{xy}=\frac{2i}{g_e^2}E_{xy}-\frac{\theta}{2\pi}~,\\
&\partial_\tau J_\tau ^{xy}=0 ~,\quad \d_x\d_yJ_\tau ^{xy}=0~.
\fe
The operators charged under this symmetry are the following Wilson strips:
\ie
&W_x(y_1,y_2)=\exp\left(\int_{y_1}^{y_2}dy\oint dx\, A_{xy}\right)~,
\\
&W_y(x_1,x_2)=\exp\left(\int_{x_1}^{x_2}dx\oint dy\, A_{xy}\right)~.
\fe
We will show below that this electric tensor symmetry has a mixed 't Hooft anomaly with the time-reversal symmetry at $\theta=\pi$.

We can couple the electric tensor symmetry to a $U(1)$ background tensor gauge field $B^{xy}_\tau$. The gauge transformation acts  as
\ie
&A_\tau \sim A_\tau + \lambda_\tau \,,~~~~A_{xy}\sim A_{xy} +\lambda_{xy}\,,\\
&B_\tau ^{xy} \sim B_\tau ^{xy}+\d_\tau \lambda_{xy}-\d_x\d_y\lambda_\tau ~.
\fe
The gauge parameters $(\lambda_\tau ,\lambda_{xy})$ are tensor gauge fields themselves with gauge symmetry
\ie
(\lambda_\tau ,\lambda_{xy}) \sim (\lambda_\tau +\d_\tau \gamma,\lambda_{xy}+\d_x\d_y\gamma)~.
\fe
The Lagrangian that couples to the background gauge field is
\ie \label{Eq:Tlagg2}
{\mathcal L}^{(A)}_{2+1}[B_\tau^{xy}]=\frac{1}{g_e^2}(E_{xy}-B_\tau ^{xy})^2+\frac{i\theta}{2\pi}(E_{xy}-B_\tau ^{xy})~.
\fe
We now examine the action of time-reversal symmetry at $\theta=0,\pi$. Time-reversal symmetry acts on the background gauge field according to $B_\tau^{xy}\rightarrow -B_\tau^{xy}$, whereas $E_{xy} \rightarrow - E_{xy}$. At $\theta=0$, this leaves both (\ref{Eq:Tlagg1}) and (\ref{Eq:Tlagg2})  invariant; hence the total theory is time-reversal symmetric.   However, at $\theta=\pi$, under time-reversal symmetry the Lagrangian transforms as
\ie\label{eq:anomalytime-rever}
\mathcal{L}^{(A)}_{2+1}[B_\tau^{xy}]\rightarrow \mathcal{L}_{2+1}^{(A)}[B_\tau^{xy}] + i B_\tau ^{xy}~.
\fe
Here we have dropped the term $i E_{xy}$ since its integral is always $2\pi i\mathbb{Z}$. In contrast, the integral of $i B_\tau^{xy}$ is generally not $2\pi i\mathbb{Z}$. Hence our theory exhibits a mixed 't Hooft anomaly between the time-reversal and the $U(1)$ electric tensor symmetry at $\theta=\pi$.

\subsection{3+1d SSPT}
We can restore the time-reversal symmetry at $\theta=\pi$ by coupling the system to a 3+1d SSPT. Denote the radial bulk coordinate by $z\geq0$. We will place our theory at the $z=0$ boundary.

The 3+1d SSPT is protected by time-reversal symmetry and a $U(1)$ tensor symmetry generated by the currents $(J_\tau^{xy},J_{[\tau z]},J_z^{xy})$ that obey
\ie
&\d_\tau J_\tau^{xy}+\d_z J_z^{xy}=0~,
\\
&\d_\tau J_{[\tau z]}+\d_x\d_y J_z^{xy}=0~,
\\
&\d_z J_{[\tau z]}=\d_x\d_y J_\tau^{xy}~.
\fe
We can couple the SSPT to a background tensor gauge field
\ie
(B_\tau ^{xy},B_{[\tau z]},B_z^{xy}) \sim (B_\tau ^{xy}+\d_\tau \lambda_{xy}-\d_x\d_y\lambda_\tau ,B_{[\tau z]}+\d_\tau \lambda_z-\d_z\lambda_\tau ,B_z^{xy}+\d_z\lambda_{xy}-\d_x\d_y\lambda_z)~.
\fe
The gauge parameters $(\lambda_\tau ,\lambda_{xy},\lambda_z)$ are gauge fields themselves with gauge transformations
\ie
(\lambda_\tau ,\lambda_{xy},\lambda_z) \sim (\lambda_\tau +\d_\tau \gamma,\lambda_{xy}+\d_x\d_y\gamma,\lambda_z+\d_z\gamma)~.
\fe
The 3+1d SSPT can be described by the classical Euclidean Lagrangian
\ie
\mathcal{L}_{3+1}=\frac{i}{2}(\d_\tau B_z^{xy}-\d_zB_\tau ^{xy}+\d_x\d_yB_{[\tau z]})~.
\fe
On a closed manifold, the theory is invariant under the time-reversal symmetry transformation
\ie
(\tau, x,y)\rightarrow (-\tau,x,y)~,\quad(B_\tau ^{xy},B_{[\tau z]},B_z^{xy})\rightarrow (-B_\tau ^{xy},-B_{[\tau z]},B_z^{xy}) \ .
\fe
 Here we have used the quantization of the following fluxes: 
\ie
\oint d\tau dx dy dz\, (\d_\tau B_z^{xy}-\d_zB_\tau ^{xy}+\d_x\d_yB_{[\tau z]})\in 2\pi\mathbb{Z}~.
\fe  
For example, a gauge field configuration that carries nontrivial flux is
\ie
B^{xy}_\tau=\frac{2\pi z}{\ell_\tau\ell_z}\left[\frac{1}{\ell_x}\delta(y-y_0)+\frac{1}{\ell_y}\delta(x-x_0)-\frac{1}{\ell_x\ell_y}\right]~,\quad B_{[\tau z]}=0~, \quad B^{xy}_z = 0~.
\fe
This is a valid configuration since $B^{xy}_\tau(z+\ell_z)-B^{xy}_\tau(z)=\d_\tau \lambda_{xy}$ with $\lambda_{xy}$ the configuration in \eqref{eq:2+1d_gauge_flux}. 
On a manifold with boundary at $z=0$, the SSPT is time-reversal invariant up to a boundary term  
\ie
S_{3+1}\rightarrow -S_{3+1}=S_{3+1d}-i \int_{z=0} d\tau dx dy\, B_\tau ^{xy}~,
\fe
which cancels the anomalous time-reversal transformation \eqref{eq:anomalytime-rever} of the 2+1d $U(1)$ tensor gauge theory on the boundary.

\section{Discussion and outlook}

In this work, we have discussed a number of qualitatively different examples of 't Hooft anomalies that arise in the context of subsystem symmetry, and shown how these anomalies can be canceled by a suitably chosen bulk theory in one higher dimension via the anomaly inflow mechanism.   Our examples illustrate that the diversity of possible 't Hooft anomalies in systems with subsystem global symmetry parallels that known to exist for ordinary global symmetry; hence a comparably rich classification of SSPT phases presumably exists.  Specifically, we have demonstrated that anomalies occur in systems with discrete and continuous subsystem symmetry, whose charged operators can be point-like or extended objects.

The subsystem symmetries examined in this work are all naturally associated with a certain foliation structure in space. 
More generally, there are lattice models exhibiting fractal  subsystem symmetries, i.e. the symmetry operators on fractal geometric objects.  
These include the fracton topological order of \cite{PhysRevA.83.042330} and also the fractal SPT of \cite{Devakul2019}.  
It is a challenging open question to develop a continuum framework to understand these fractal subsystem symmetries. 

Even for the examples discussed here, where the  anomalous theory has a natural foaliation structure, there can be more than one possible extension of the boundary foliation structure into the bulk.  For example, in Section \ref{sec:2dZN}, we  encountered an example of  a $\mathbb{Z}_N$ subsystem symmetry anomaly in 1+1d that can be canceled by two distinct 2+1d SSPTs with different bulk foliation structures. 
Thus the correspondence between anomalies and SPTs appears to be more subtle than for  ordinary global symmetries, where anomalies can be classified by SPTs in one dimension higher.  
To achieve a classification of subsystem anomalies via SSPTs in one dimension higher, one needs to first understand the possible extensions of the boundary foliation structure into the bulk.\footnote{The foliation structure of a theory with subsystem symmetries is somewhat analogous to the tangential structure (e.g., the spin structure) in the discussion of ordinary global symmetries.  For example, various physical observables, such as the ground state degeneracy, depend not only on the geometry, but also on the choice of a foliation in systems with subsystem symmetries \cite{Slagle:2017mzz,Shirley:2017suz}. This is analogous to the dependence on the choice of the spin structure in a fermionic theory.  In the ordinary correspondence between the boundary anomaly and the bulk SPT, one assumes the tangential structure is \textit{stable} in the sense that it can be defined in all spacetime dimensions. It would be interesting to understand the corresponding mathematical structure for foliated manifolds. We thank Kantaro Ohmori for discussions on this point. }

The perspective adopted in this paper is to start from a boundary system with a given anomaly, and then identify a bulk theory to cancel this anomaly.  
Alternatively, one can start with a bulk SSPT, and analyze the anomaly inflow into the  boundary. 
Since our SSPT does not have continuous spatial rotation symmetry, the anomaly inflow depends sensitively on the choice of the boundary. 
For example, the  anomalies might be different for the boundaries along different directions in space. 
This possibility also makes the correspondence between the boundary anomaly and the bulk SSPT more intricate.

We leave a systematic investigation of these questions for future studies.

\section*{Acknowledgements}

We thank N.\ Seiberg for suggesting this topic to us and for initial collaborations in this project. 
We also thank Z.\ Komargodski and K.\ Ohmori  for discussions.
FJB is grateful for the financial support of the Carnegie corporation of New York, and of the National Science foundation NSF DMR 1928166. 
PG was supported by the Physics Department of Princeton University. HTL was supported in part by a Centennial Fellowship and a Charlotte Elizabeth Procter Fellowship from Princeton University, a Croucher fellowship  from the Croucher Foundation, the Packard Foundation, the Physics Department of Princeton University and the Center for Theoretical Physics at MIT. 
SHS was supported by the Simons Collaboration on Ultra-Quantum Matter, which is a grant from the Simons Foundation (651440, NS). The authors of this paper were ordered alphabetically.

\appendix
\section{Anomalies in ordinary quantum field theories}\label{app-A}
In this appendix, we will review several well-known examples of systems with 't Hooft anomalies of ordinary global symmetries, and the associated symmetry protected topological (SPT) phases that cancel the anomalies via anomaly inflow.

\subsection{Anomalies of $U(1)$ symmetries}\label{app:U1U1}

\subsubsection{$U(1)\times U(1)$ symmetry: 1+1d compact boson}\label{app:anom_compact boson}
Consider the 1+1d compact boson theory described by the Euclidean Lagrangian
\ie\label{app-compactboson-Lag}
\mathcal L_{1+1}^{(\phi)} = \frac{\beta}{2} \left[ (\partial_\tau \phi)^2 + (\partial_x \phi)^2 \right]~,
\fe
where $\phi$ is a compact scalar, i.e., $\phi \sim \phi + 2\pi$. In the path integral, we sum over winding configurations of $\phi$. On a torus of size $\ell_\tau, \ell_x$, they obey the boundary conditions
\ie\label{app-compactboson-boundary_condition}
&\phi(\tau+\ell_\tau,x)=\phi(\tau,x)+2\pi w_\tau~,\qquad w_\tau\in\mathbb{Z}~,
\\
&\phi(\tau,x+\ell_x)=\phi(\tau,x)+2\pi w_x~,\qquad w_x\in\mathbb{Z}~.
\fe 

The Lagrangian \eqref{app-compactboson-Lag} is invariant under the $U(1)$ momentum global symmetry that shifts
\ie
\phi(\tau,x) \rightarrow \phi(\tau,x) + c~,
\fe
where $c$ is a constant. The Noether current for this symmetry are
\ie
&J_\tau = i\beta \partial_\tau \phi~,\qquad J_x = i\beta \partial_x \phi~,
\\
&\partial_\tau J_\tau + \partial_x J_x=0~.
\fe
There is also a $U(1)$ winding symmetry with Noether current
\ie
&\tilde J_\tau = \frac{1}{2\pi} \partial_x \phi~, \qquad \tilde J_x = -\frac{1}{2\pi} \partial_\tau \phi~,
\\
&\partial_\tau \tilde J_\tau + \partial_x \tilde J_x=0~.
\fe
A field configuration that carries a nontrivial winding charge is $\phi = 2\pi x/\ell_x$, which satisfies the boundary condition \eqref{app-compactboson-boundary_condition}.

We can couple the theory to the background gauge fields $A=(A_\tau,A_x)$ and $\tilde A=(\tilde A_\tau,\tilde A_x)$ of the $U(1)$ momentum and the $U(1)$ winding symmetry.
The Lagrangian after coupling is
\ie
\mathcal L^{(\phi)}_{1+1}[A,\tilde A] = \frac{\beta}{2} \left[ (\partial_\tau \phi - A_\tau)^2 + (\partial_x \phi - A_x)^2 \right] + \frac{i}{2\pi}\left[ \tilde A_\tau (\partial_x \phi - A_x) - \tilde A_x (\partial_\tau \phi - A_\tau) \right]~.
\fe
It is not invariant under the gauge symmetry:
\ie\label{app-compactboson-gaugesym}
\phi\sim\phi+\alpha~, \qquad A \sim A + d\alpha~, \qquad \tilde A \sim \tilde A + d\tilde \alpha~.
\fe
Instead, it is shifted by 
\ie\label{app-compactboson-anom}
\mathcal L^{(\phi)}_{1+1}[A,\tilde A] \rightarrow \mathcal L^{(\phi)}_{1+1}[A,\tilde A] + \frac{i}{2\pi} \tilde \alpha (\partial_\tau A_x - \partial_x A_\tau)~.
\fe
This signals an 't Hooft anomaly of the $U(1)\times U(1)$ global symmetry.

Consider a 2+1d SPT protected by a $U(1)\times U(1)$ global symmetry described by the classical Euclidean Lagrangian\footnote{In this appendix, we will often expand the differential forms in components to compare with the analogous expressions for the subsystem symmetries in the main texts. We will often ignore the the volume form, such as $d\tau\wedge dx\wedge dy$, in these expressions.}
\ie\label{app-compactboson-spt}
\mathcal L_{2+1}[A,\tilde A] &= \frac{i}{2\pi} \tilde A dA 
\\
&= \frac{i}{2\pi}\left[ \tilde A_\tau (\partial_x A_y - \partial_y A_x) - \tilde A_x (\partial_\tau A_y - \partial_y A_\tau) + \tilde A_y (\partial_\tau A_x - \partial_x A_\tau) \right]
~,
\fe
where $A = (A_\tau,A_x,A_y)$ and $\tilde A = (\tilde A_\tau,\tilde A_x,\tilde A_y)$ are background $U(1)\times U(1)$ gauge fields with gauge symmetry,
\ie
A \sim A+d\alpha~,\qquad \tilde A \sim \tilde A + d \tilde \alpha~.
\fe
The Lagrangian \eqref{app-compactboson-spt} is invariant under the gauge symmetry up to a total derivative: 
\ie
\frac{i}{2\pi} d(\tilde \alpha dA) = \frac{i}{2\pi} \left[ \partial_\tau\big(\tilde \alpha (\partial_x A_y - \partial_y A_x)\big) - \partial_x \big(\tilde \alpha (\partial_\tau A_y - \partial_y A_\tau)\big) + \partial_y \big(\tilde \alpha (\partial_\tau A_x - \partial_x A_\tau)\big) \right]
~.
\fe
On a manifold with boundary at $y=0$, the anomaly inflow cancels the anomaly \eqref{app-compactboson-anom}.

\subsubsection{$U(1)$ symmetry: 1+1d compact chiral boson}\label{app:chiral_boson}

Consider the 1+1d compact chiral boson   described by the Euclidean action
\ie\label{app-chiralboson-Lag}
S_{1+1}^{(\varphi)} = \frac{iN}{4\pi}\oint d\tau dx \,\d_\tau\varphi\d_x\varphi~,
\fe
where $N\in\mathbb{Z}$ and $\varphi$ is a compact scalar i.e. $\varphi \sim \varphi+2\pi$. 
More generally, one can add a velocity term $(\partial_x\varphi)^2$ to the Lagrangian, but we will consider the special case when this term is absent.

On a torus of size $\ell_\tau,\ell_x$, we sum over the winding configurations that obey the boundary conditions\footnote{More generally, in the path integral, we can include configurations that wind in the time direction. But using the gauge symmetry \eqref{app-chiralboson-gauge}, we can always shift a winding configuration $\varphi(\tau,x)$ by a $g(\tau)$ with opposite winding such that in the path integral we can restrict to configurations that do not wind in the time direction.}
\ie\label{app-chiralboson-boundary_condition}
&\varphi(\tau+\ell_\tau,x)=\varphi(\tau,x)~,
\\
&\varphi(\tau,x+\ell_x)=\varphi(\tau,x)+2\pi w_x~,\qquad w_x\in\mathbb{Z}~,
\fe 
in the path integral. The theory has a gauge symmetry
\ie\label{app-chiralboson-gauge}
\varphi(\tau,x)\sim\varphi(\tau,x)+g(\tau)~,
\fe
where $g(\tau +\ell_\tau)=g(\tau)$.

The theory has a $\text{L}U(1)/U(1)$ momentum global symmetry that shifts \cite{Moore:1989yh,Elitzur:1989nr}
\ie\label{LU1}
\varphi(\tau,x) \rightarrow \varphi(\tau,x) +f(x)~,
\fe
where $f(x)$ is a map from $S^1$ to $U(1)$. In particular, $f(x)$ can wind with $f(x+\ell_x)=f(x)+2\pi\mathbb{Z}$. The zero mode of $f(x)$ is not a global symmetry but a gauge symmetry included in \eqref{app-chiralboson-gauge}. Because of this, the global symmetry has a $U(1)$ quotient. The symmetry is generated by the current
\ie
&J_\tau=-\frac{N}{2\pi}\d_x\varphi~,
\\
&\d_\tau J_\tau=0~.
\fe
Although a constant shift of $\varphi$ is a gauge symmetry, the zero-mode charge $Q=\oint dx\, J_\tau$ can be non-trivial since the winding configuration $\varphi = 2\pi x/\ell_x$ carries a nontrivial charge $Q=-N$.

We can couple the current $J_\tau$ to a $U(1)$ background gauge field $A=(A_\tau, A_x)$ \cite{Sonnenschein:1988ug}
\ie
{S}^{(\varphi)}_{1+1}[A_\tau,A_{x}]=\frac{iN}{4\pi}\oint d\tau dx\left(\d_\tau\varphi\d_x\varphi-2 A_\tau\d_x\varphi+ A_\tau A_{x}\right)~.
\fe
Note that since the current only has the $J_\tau$ component, the background gauge field $A_x$ is not coupled to any current. Here we include in the Lagrangian a classical counterterm $\frac{iN}{4\pi} A_\tau A_{x}$ for later convenience. This does not affect the 't Hooft anomaly.
The Lagrangian is not invariant under the gauge symmetry,
\ie
\varphi\sim \varphi +\alpha~,\quad A\sim A+d\alpha~.
\fe
Rather, it is shifted by
\ie\label{app-chiralboson-anom}
S^{(\varphi)}_{1+1}[A_\tau,A_{x}] \to S^{(\varphi)}_{1+1}[A_\tau,A_{x}]-\frac{iN}{4\pi}\oint d\tau dx\, \alpha (\d_\tau A_{x}-\d_{x}A_\tau)~.
\fe
It signals an 't Hooft anomaly.

This anomaly can be canceled by coupling the system to a 2+1d $U(1)$ SPT described by the classical Euclidean Lagrangian
\ie
\mathcal{L}_{2+1}[A_\tau,A_{x},A_y]
&=
-\frac{iN}{4\pi}AdA
\\
&=
-\frac{iN}{4\pi}\big[A_\tau(\d_x A_{y}-\d_y A_x)+A_x(\d_y A_{\tau}-\d_\tau A_y)+A_{y}(\d_\tau A_x-\d_x A_{\tau})\big]~.
\fe
The Lagrangian is invariant under the gauge symmetry up to a total derivative 
\ie
-\frac{iN}{4\pi} d(\alpha dA) = -\frac{iN}{4\pi} \left[ \partial_\tau\big( \alpha (\partial_x A_y - \partial_y A_x)\big) - \partial_x \big( \alpha (\partial_\tau A_y - \partial_y A_\tau)\big) + \partial_y \big( \alpha (\partial_\tau A_x - \partial_x A_\tau)\big) \right]~.
\fe
On a manifold with a $y=0$ boundary, the anomaly inflow cancels the anomaly \eqref{app-chiralboson-anom}.

\subsection{Anomalies of $\mathbb{Z}_N$ symmetries}

\subsubsection{$\mathbb{Z}_N\times\mathbb{Z}_N$ one-form symmetry: 2+1d $\mathbb{Z}_N$ gauge theory}\label{app:2+1d_ZN_gauge_theory}
The Euclidean Lagrangian of the 2+1d $\mathbb Z_N$ gauge theory is \cite{Maldacena:2001ss,Banks:2010zn,Kapustin:2014gua}
\ie\label{app-ZN-Lag}
\mathcal L_{2+1} = \frac{iN}{2\pi} \tilde A dA~,
\fe
where $A = (A_\tau,A_x,A_y)$ and $\tilde A = (\tilde A_\tau,\tilde A_x,\tilde A_y)$ are $U(1)$ gauge fields with gauge symmetry
\ie\label{app-ZN-gaugesym}
A \sim A + d\alpha~, \qquad \tilde A \sim \tilde A + d\tilde \alpha~.
\fe
Their equations of motion constrain each other to be $\mathbb{Z}_N$ gauge fields. This theory describes the low energy physics of the $\mathbb Z_N$ toric code.

The theory \eqref{app-ZN-Lag} has an electric $\mathbb Z_N$ one-form global symmetry, and a magnetic $\mathbb Z_N$ one-form global symmetry. The electric one-form symmetry is generated by the Wilson loops of $\tilde A$, $\tilde W(\tilde C) = \exp( i \oint_{\tilde C} \tilde A)$, and the charged operators are the Wilson loops of $A$, $W(C) = \exp( i \oint_C A)$. On the other hand, the magnetic one-form symmetry is generated by the Wilson loops of $A$, and the charged operators are the Wilson loops of $\tilde A$.

These operators satisfy
\ie\label{eq:2+1d_ZN_gauge_theory_operator_algebra}
W(C)^N = \tilde W(\tilde C)^N = 1~,\qquad W(C)^n \tilde W(\tilde C)^m = e^{\frac{2\pi i}{N}n m I(C,\tilde C)} \tilde W(\tilde C)^m W(C)^n~,
\fe
where $C$ and $\tilde C$ are restricted to curves in space, and $I(C,\tilde C)$ is their intersection number. The fact that the two symmetry operators are charged under each other signals a mixed 't Hooft anomaly between the two $\mathbb Z_N$ one-form symmetries \cite{Gaiotto:2014kfa,Gomis:2017ixy,Hsin:2018vcg,Wen:2018zux}. 
One consequence of this anomaly is that, on a torus, the $\mathbb{Z}_N$ gauge theory has $N^2$ degenerate ground states that live in the minimal representation of the operator algebra  \eqref{eq:2+1d_ZN_gauge_theory_operator_algebra}. 

The Wilson lines $W$ and $\tilde W$ are the worldlines of the electric and magnetic
anyons in the microscopic toric code. The nontrivial braiding between the electric and the
magnetic anyons becomes the 't Hooft anomaly of the emergent $\mathbb{Z}_N\times\mathbb{Z}_N$ one-form global symmetry.

The anomaly can also be detected by coupling the theory \eqref{app-ZN-Lag} to background two-form gauge fields $B$ and $\tilde B$ of the $\mathbb Z_N \times \mathbb Z_N$ one-form global symmetry:
\ie\label{app-ZN-Lag-bkg-gauge}
\mathcal L_{2+1}[B,\tilde B] = \frac{iN}{2\pi} \left[ \tilde A (dA - B) - A \tilde B \right] + \frac{iN}{2\pi} \tilde \phi dB + \frac{iN}{2\pi} \phi d\tilde B~,
\fe
where $B$ and $\tilde B$ are classical $U(1)$ two-form gauge fields, and $\tilde \phi$ and $\phi$ are dynamical compact scalar fields that constrain $B$ and $\tilde B$ to be $\mathbb Z_N$ two-form gauge fields. The dynamical gauge symmetry \eqref{app-ZN-gaugesym} acts on $\phi$ and $\tilde \phi$ as
\ie
\phi \sim \phi - \alpha~, \qquad \tilde \phi \sim \tilde \phi - \tilde \alpha~.
\fe
The background one-form gauge symmetry acts as
\ie\label{app-ZN-bkg-gaugesym}
&A \sim A + \lambda~,\qquad B \sim B + d\lambda~,
\\
&\tilde A \sim \tilde A + \tilde \lambda~, \qquad \tilde B \sim \tilde B + d\tilde \lambda~,
\fe
where the one-form gauge parameters $\lambda$ and $\tilde \lambda$ have their own gauge symmetries
\ie
\lambda \sim \lambda + d\gamma~, \qquad \tilde \lambda \sim \tilde \lambda + d\tilde \gamma~.
\fe
The Lagrangian \eqref{app-ZN-Lag-bkg-gauge} is not invariant under the gauge symmetry \eqref{app-ZN-bkg-gaugesym}:
\ie\label{app-ZN-anom}
\mathcal L_{2+1}[B,\tilde B] \rightarrow \mathcal L_{2+1}[B,\tilde B] - \frac{iN}{2\pi} \left[ \tilde \lambda B + \lambda (\tilde B + d\tilde \lambda) \right]~.
\fe
This signals an 't Hooft anomaly of the $\mathbb Z_N \times \mathbb Z_N$ one-form symmetry.

Consider a 3+1d SPT protected by the $\mathbb Z_N \times \mathbb Z_N$ one-form symmetry described by the Lagrangian \cite{Thorngren:2015gtw,Hsin:2018vcg}
\ie\label{app-ZN-spt}
\mathcal L_{3+1}[B,\tilde B] = \frac{iN}{2\pi} \tilde C dB + \frac{iN}{2\pi} C d\tilde B - \frac{iN}{2\pi} B \tilde B~,
\fe
where $B$ and $\tilde B$ are $U(1)$ two-form background gauge fields, and $C$ and $\tilde C$ are dynamical one-form gauge fields that constrain $\tilde B$ and $B$ to be $\mathbb Z_N$ two-form gauge fields. The gauge symmetry is
\ie\label{app-ZN-spt-gaugesym}
&C \sim C + d\chi + \lambda~, \qquad B \sim B + d\lambda~,
\\
&\tilde C \sim \tilde C + d\tilde \chi + \tilde \lambda~, \qquad \tilde B \sim \tilde B + d\tilde \lambda~,
\fe
where the background one-form gauge parameters $\lambda$ and $\tilde \lambda$ have their own gauge symmetries
\ie
\lambda \sim \lambda + d\gamma~, \qquad \tilde \lambda \sim \tilde \lambda + d\tilde \gamma~.
\fe
The 3+1d SPT \eqref{app-ZN-spt} is the low-energy limit of a Walker-Wang model \cite{Walker:2011mda}.

The Lagrangian \eqref{app-ZN-spt} is invariant under the background gauge symmetry \eqref{app-ZN-spt-gaugesym} up to total derivatives
\ie
\mathcal L_{3+1}[B,\tilde B] \rightarrow \mathcal L_{3+1}[B,\tilde B] - \frac{iN}{2\pi} d\left[ \tilde \lambda B + \lambda (\tilde B + d\tilde \lambda)  \right]~.
\fe
On a manifold with boundary at $z=0$, the anomaly inflow cancels the anomaly \eqref{app-ZN-anom}.

\subsubsection{$\mathbb{Z}_N\times\mathbb{Z}_N$ symmetry: quantum mechanics of $N$ degenerate ground states}\label{app:1+1d_pqdot}

Consider the quantum mechanics of $N$ degenerate ground states. When the Euclidean time is noncompact, the theory is described by the Euclidean action
\ie\label{app-pdotq-action_naive}
S_{0+1} = \frac{iN}{2\pi} \int d \tau~  p(\tau) \dot q(\tau)~,
\fe
where $N\in\mathbb{Z}$ and $p(\tau),q(\tau)$ are both circle-valued fields,
\ie\label{app-pdotq-ident}
p(\tau) \sim p(\tau) + 2\pi~,\qquad q(\tau) \sim q(\tau)+2\pi~.
\fe
If the time is compact with a periodicity $\ell_\tau$, the fields $p(\tau),q(\tau)$ can wind around the Euclidean time circle
\ie
&p(\tau + \ell_\tau) = p(\tau) + 2\pi n_p~,\qquad n_p \in \mathbb Z~,
\\
&q(\tau + \ell_\tau) = q(\tau) + 2\pi n_q~,\qquad n_q \in \mathbb Z~,
\fe
and all values of $n_p,n_q$ are summed over in the path integral. As a result, the action \eqref{app-pdotq-action_naive} is no longer well-defined unless it is supplemented with a correction term\cite{Bauer:2004nh,Cordova:2019jnf,Cordova:2019uob,Gorantla:2021svj}:
\ie\label{app-pdotq-action}
S_{0+1} = \frac{iN}{2\pi} \int_{\tau_*}^{\tau_* + \ell_\tau} d \tau~  p(\tau) \dot q(\tau) - iN n_p q(\tau_*)~.
\fe
It is straightforward to see that the action \eqref{app-pdotq-action} is independent of the reference Euclidean time $\tau_*$, and is invariant (modulo $2\pi i \mathbb Z$) under the identifications \eqref{app-pdotq-ident}.

The action \eqref{app-pdotq-action} has a $\mathbb Z_N \times \mathbb Z_N$ global symmetry that shifts 
\ie
&p(\tau) \rightarrow p(\tau) + \frac{2\pi m_p}{N}~, \qquad m_p \in \mathbb Z~,
\\
&q(\tau) \rightarrow q(\tau) + \frac{2\pi m_q}{N}~, \qquad m_q \in \mathbb Z~.
\fe
It is generated by $U = e^{iq}$ and $V=e^{ip}$ which satisfy 
\ie\label{app-pdotq-commrel}
U^N = V^N = 1~,\qquad U V = e^\frac{2\pi i}{N} VU~.
\fe
Quantizing the action leads to a Hilbert space of $N$ degenerated ground states, which are in the minimal representation of this algebra. The nontrivial commutation relations in \eqref{app-pdotq-commrel} mean that the Hilbert space transforms projectively under the $\mathbb Z_N \times \mathbb Z_N$ symmetry. This implies an 't Hooft anomaly of the symmetry.

The anomaly can also be detected by coupling the system to $\mathbb Z_N \times \mathbb Z_N$ background gauge fields:
\ie\label{app-pdotq-bkg-gauge}
S_{0+1}[A_\tau,\tilde A_\tau] = \frac{iN}{2\pi} \oint d \tau \left[p (\dot q - A_\tau) + \tilde A_\tau q \right]~.
\fe
Here, we omit the correction terms, and restrict the holonomies of the $U(1)$ background gauge fields $A_\tau$ and $\tilde A_\tau$ to be $\mathbb{Z}_N$-valued.\footnote{We can impose these restrictions by adding to the action \eqref{app-pdotq-bkg-gauge} the following terms:
	\ie
	\frac{i}{2\pi}\oint d\tau \left[ \tilde \chi(\dot \psi - N A_\tau) + \chi (\dot{\tilde \psi} - N \tilde A_\tau) \right]~,
	\fe
	where $\chi,\tilde \chi$ are real-valued Lagrange multipliers, and $\psi,\tilde \psi$ are circle-valued. The latter fields transform under the background gauge symmetry \eqref{app-pdotq-gauge-sym} as
	\ie
	\psi \sim \psi + N \alpha~,\qquad \tilde \psi \sim \tilde \psi + N\tilde \alpha~.
	\fe
	Since these terms do not affect the anomaly, we omit them in \eqref{app-pdotq-bkg-gauge}.}
The background gauge symmetry acts as
\ie\label{app-pdotq-gauge-sym}
&q \sim q + \alpha~, \qquad A_\tau \sim A_\tau + \dot \alpha~,
\\
&p \sim p + \tilde \alpha~, \qquad \tilde A_\tau \sim \tilde A_\tau + \dot{\tilde \alpha}~.
\fe
The action \eqref{app-pdotq-bkg-gauge} is not invariant under the gauge symmetry \eqref{app-pdotq-gauge-sym},
\ie\label{app-pdotq-anom}
S_{0+1}[A_\tau,\tilde A_\tau] \rightarrow S_{0+1}[A_\tau,\tilde A_\tau] + \frac{iN}{2\pi} \oint d\tau ~( \alpha \tilde A_\tau - \tilde \alpha A_\tau - \tilde \alpha \dot \alpha )~,
\fe
which signals an 't Hooft anomaly of the $\mathbb Z_N \times \mathbb Z_N$ global symmetry.

Consider a 1+1d SPT protected by a $\mathbb Z_N \times \mathbb Z_N$ global symmetry. The SPT is described by the Euclidean Lagrangian
\ie\label{app-pdotq-spt}
\mathcal{L}_{1+1}[A,\tilde A] &=  \frac{iN}{2\pi} \left[\tilde \Phi dA + \Phi d \tilde A - A \tilde A \right]
\\
& = \frac{iN}{2\pi} \left[ \tilde \Phi (\partial_\tau A_x - \partial_x A_\tau) + \Phi (\partial_\tau \tilde A_x - \partial_x \tilde A_\tau) - (A_\tau \tilde A_x - A_x \tilde A_\tau) \right]~,
\fe
where $\tilde \Phi$ and $\Phi$ are dynamical compact scalar fields that constrain the $U(1)$ background gauge fields $A=(A_\tau,A_x)$ and $\tilde A = (\tilde A_\tau,\tilde A_x)$ to be $\mathbb Z_N$-valued. The gauge symmetry acts as
\ie
&A \sim A + d\alpha~, \qquad \Phi \sim \Phi - \alpha~,
\\
&\tilde A \sim \tilde A + d \tilde \alpha~, \qquad \tilde \Phi \sim \tilde \Phi + \tilde \alpha~.
\fe
The Lagrangian \eqref{app-pdotq-spt} is invariant under the gauge symmetry up to total derivatives:
\ie
-\frac{iN}{2\pi}  d(\alpha \tilde A - \tilde \alpha A - \tilde \alpha d \alpha)
=-\frac{iN}{2\pi} \left[ \partial_\tau(\alpha \tilde A_x - \tilde \alpha A_x - \tilde \alpha \partial_x \alpha) -  \partial_x(\alpha \tilde A_\tau - \tilde \alpha A_\tau - \tilde \alpha \partial_\tau \alpha) \right]~.
\fe
On a manifold with boundary at $x=0$, the anomaly inflow cancels the anomaly \eqref{app-pdotq-anom}.

\subsection{Time-reversal and $U(1)$ one-form symmetry: 1+1d $U(1)$ gauge theory}\label{app:1+1d_U(1)_gauge_theory}

Consider the 1+1d $U(1)$ Maxwell gauge theory with a $\theta$-term,
\ie\label{app-maxwell-Lag}
\mathcal L_{1+1}^\text{Max} = \frac{1}{g^2} E_x^2 + \frac{i\theta}{2\pi} E_x~,
\fe
where $E_x = \partial_\tau  A_x - \partial_x A_\tau $ is the electric field. 
The one-form gauge field $A=(A_\tau ,A_x)$ has the ordinary  gauge symmetry:
\ie
A \sim A + d\alpha~.
\fe
The $\theta$-parameter has $2\pi$ periodicity because of flux quantization $\oint d\tau dx\, E_x \in 2\pi \mathbb Z$. 

The theory \eqref{app-maxwell-Lag} has a $U(1)$ electric one-form symmetry with Noether current
\ie
&J_{[\tau x]} = \frac{2i}{g^2} E_x - \frac{\theta}{2\pi}~,
\\
&\partial_\tau  J_{[\tau x]} = 0~, \qquad \partial_x J_{[\tau x]} = 0~.
\fe
At $\theta = 0,\pi$, the theory also has a time-reversal symmetry:
\ie
(\tau, x)\rightarrow (-\tau,x)~,\quad (A_\tau,A_{x})\rightarrow(-A_\tau,A_{x})~.
\fe

We can couple \eqref{app-maxwell-Lag} to the background two-form gauge field $B = B_{[\tau x]}$ of the electric one-form symmetry:
\ie
\mathcal{L}_{1+1}^\text{Max}[B] =  \frac{1}{g^2} (E_x - B_{[\tau x]})^2 + \frac{i\theta}{2\pi} (E_x - B_{[\tau x]}) ~.
\fe
The one-form background gauge symmetry is
\ie
A \sim A + \lambda~,\qquad B \sim B + d\lambda~,
\fe
where the one-form gauge parameter $\lambda = (\lambda_\tau ,\lambda_x)$ itself has a gauge symmetry,
\ie
\lambda \sim \lambda + d\gamma~.
\fe
With a nontrivial $B_{[\tau x]}$, the time-reversal symmetry is unbroken at $\theta = 0$ but it is broken at $\theta = \pi$ with the anomalous transformation
\ie\label{app-maxwell-anom}
\mathcal{L}_{1+1}^\text{Max}[B] \rightarrow \mathcal{L}_{1+1}^\text{Max}[B] + i B_{[\tau x]}~.
\fe
This signals a mixed 't Hooft anomaly between the electric one-form symmetry and the time reversal symmetry at $\theta = \pi$ \cite{Gaiotto:2017yup,Komargodski:2017dmc}.

We can restore the time-reversal symmetry at $\theta = \pi$ by coupling the system to a 2+1d SPT protected by the time-reversal and the $U(1)$ one-form symmetry. 
For simplicity, we will assume the bulk manifold is orientable and refer the readers to 
 \cite{Komargodski:2017dmc} for a more general discussion.

In the bulk, the $U(1)$ one-form symmetry is coupled to a  two-form background gauge field $B=(B_{[\tau x]},B_{[\tau y]},B_{[xy]})$ with gauge symmetry
\ie
B \sim B + d\lambda~,
\fe
where the one-form gauge parameter $\lambda = (\lambda_\tau ,\lambda_x,\lambda_y)$ itself has a gauge symmetry
\ie
\lambda \sim \lambda + d\gamma~.
\fe
The 2+1d SPT is described by the Euclidean Lagrangian 
\ie\label{app-maxwell-spt}
\mathcal{L}_{2+1}[B] = - \frac i2 dB = -\frac i2(\partial_\tau  B_{[xy]} - \partial_x B_{[\tau y]} + \partial_y B_{[\tau x]})~.
\fe
On a closed orientable manifold, the SPT is time-reversal invariant because of flux quantization $\oint dB \in 2\pi \mathbb Z$. On a manifold with boundary at $y=0$, it is time-reversal invariant up to a boundary term: 
\ie
S_{2+1}[B] \rightarrow -S_{2+1}[B] = S_{2+1}[B] - i\int_{y=0} d\tau dx~ B_{[\tau x]}~,
\fe
which cancels the anomalous time-reversal transformation \eqref{app-maxwell-anom} of the 1+1d $U(1)$ Maxwell theory on the boundary.

\bibliographystyle{JHEP}
\bibliography{bib}

\end{document}